\documentclass[sigconf, nonacm]{acmart}

\AtBeginDocument{%
  \providecommand\BibTeX{{%
    \normalfont B\kern-0.5em{\scshape i\kern-0.25em b}\kern-0.8em\TeX}}}

\setcopyright{acmcopyright}
\copyrightyear{2021}
\acmYear{2021}
\acmBooktitle{Arxiv}

\usepackage[nolist]{acronym}
\usepackage{multirow}
\usepackage{multicol}
\usepackage{subcaption}
\usepackage{listings}
\newenvironment{conditions*}
  {\par\vspace{\abovedisplayskip}\noindent
   \tabularx{\columnwidth}{>{$}l<{$} @{\ : } >{\raggedright\arraybackslash}X>{$}l<{$} @{\ : } >{\raggedright\arraybackslash}X}}
  {\endtabularx\par\vspace{\belowdisplayskip}}

\newif\ifarxiv

\newif\ifopen
\newif\iffinal

\opentrue %
\finaltrue

\ifarxiv
\opentrue
\finaltrue
\fi

\makeatletter
\def\lst@makecaption{%
  \def\@captype{table}%
  \@makecaption
}
\makeatother
\usepackage{booktabs}
\usepackage{textcomp}
\usepackage[colorinlistoftodos,prependcaption]{todonotes}
\usepackage{regexpatch}
\makeatletter
\xpatchcmd{\@todo}{\setkeys{todonotes}{#1}}{\setkeys{todonotes}{inline,#1}}{}{}
\makeatother
\usepackage{xcolor}
\usepackage[binary-units, per-mode=symbol]{siunitx}
\def\BibTeX{{\rm B\kern-.05em{\sc i\kern-.025em b}\kern-.08em
    T\kern-.1667em\lower.7ex\hbox{E}\kern-.125emX}}

\iffinal
\newcommand{\colorrevi}{black}
\else
\newcommand{\colorrevi}{blue}
\fi
\newcommand{\rev}[1]{{\textcolor{\colorrevi}{#1}}}

\begin{document}

\begin{acronym}
\acro{ALM}{Adaptive Logic Modules}
\acro{BRAM}{Block RAM}
\acro{DDR}[DDR-SDRAM]{Double Data Rate Synchronous Dynamic Random Access Memory}
\acro{DSP}{Digital Signal Processor}
\acro{FF}{Flop-Flop}
\acro{HBM2}{High Bandwidth Memory 2}
\acro{PSN}[PSN]{Packet Switched Network}
\acro{CSN}{Circuit Switched Network}
\acro{MPI}{Message Passing Interface}
\acro{LUT}{Lookup Table}
\acro{FPGA}{Field Programmable Gate Array}
\acro{HPC}{High Performance Computing}
\acro{HDL}{Hardware Description Language}
\acro{HPL}{High Performance LINPACK}
\acro{HLS}{High Level Synthesis}
\acro{HPCC}{HPC Challenge}
\acro{OpenCL}{Open Computing Language}
\acro{GPU}{Graphics Processing Unit}
\acro{FLOPS}{Floating Point Operations per Second}
\acro{SDK}{Software Development Kit}
\acro{BSP}{Board Support Package}
\acro{LSU}{Load Store Unit}
\acro{SVM}{Shared Virtual Memory}
\acro{FLOP}{Floating Point Operation}
\acro{GUPS}{Giga Updates Per Second}
\acro{EP}{Embarrassingly Parallel}
\acro{RNG}{Random Number Generator}
\acro{PC$^2$}{Paderborn Center for Parallel Computing}
\acro{IEC}{Intel External Channels}
\end{acronym}

\title{Multi-FPGA Designs and Scaling of HPC Challenge Benchmarks via MPI and Circuit-Switched Inter-FPGA Networks}

\ifopen
\author{Marius Meyer}
\email{marius.meyer@uni-paderborn.de}
\affiliation{%
  \institution{Paderborn University}
  \department{Department of Computer Science and Paderborn Center for Parallel Computing (PC$^2$)}
  \streetaddress{Warburger Str. 100, 33098 Paderborn}
  \country{Germany}
}
\author{Tobias Kenter}
\email{tobias.kenter@uni-paderborn.de}
\affiliation{%
\institution{Paderborn University}
\department{Department of Computer Science and Paderborn Center for Parallel Computing (PC$^2$)}
\streetaddress{Warburger Str. 100, 33098 Paderborn}
\country{Germany}
}

\author{Christian Plessl}
\email{christian.plessl@uni-paderborn.de}
\affiliation{%
\institution{Paderborn University}
\department{Department of Computer Science and Paderborn Center for Parallel Computing (PC$^2$)}
\streetaddress{Warburger Str. 100, 33098 Paderborn}
\country{Germany}
}

\else
\author{Anonymous Author(s)}
\fi

\begin{abstract}
While FPGA accelerator boards and their respective high-level design tools are maturing, there is still a lack of multi-FPGA applications, libraries, and not least, benchmarks and reference implementations towards sustained HPC usage of these devices. 
As in the early days of GPUs in HPC, for workloads that can reasonably be decoupled into loosely coupled working sets, multi-accelerator support can be achieved by using standard communication interfaces like MPI on the host side. %
However, for performance and productivity, some applications can profit from a tighter coupling of the accelerators. 
FPGAs offer unique opportunities here when extending the dataflow characteristics to their communication interfaces.

In this work, we extend the HPCC FPGA benchmark suite by multi-FPGA support and three missing benchmarks that particularly characterize or stress inter-device communication: b\_eff, PTRANS, and LINPACK. 
With all benchmarks implemented for current boards with Intel and Xilinx FPGAs, we established a baseline for multi-FPGA performance. %
Additionally, for the communication-centric benchmarks, we explored the potential of direct FPGA-to-FPGA communication with a circuit-switched inter-FPGA network that is currently only available for one of the boards. 
The evaluation with parallel execution on up to 26 FPGA boards makes use of one of the largest academic FPGA installations.

\end{abstract}


  \maketitle

  \acresetall

\section{Introduction}
\label{sec:introduction}
The heterogeneity of \ac{HPC} systems increased over the last years, and heterogeneous systems take an important role in the Top 500 list \cite{top500}.
\acp{FPGA} are good candidates for the acceleration of \ac{HPC} systems since current \ac{HLS} tool flows drastically decreased the development time while still offering high-quality results.
With novel \ac{HPC} systems emerging, coming with a hybrid network consisting of an inter-CPU and a separate inter-\ac{FPGA} network, also new opportunities for scaling applications on these systems arise.
Handling the required communication for the accelerator workloads over communication interfaces like \ac{MPI} through the inter-CPU network and updating the data on the device via PCIe comes with limitations.
The workloads need to be reasonably decoupled to allow efficient, concurrent operation on multiple accelerators, which is not always possible.
Some applications can profit from higher bandwidths and shorter latencies achieved by tighter coupling of the accelerators using direct inter-\ac{FPGA} communication.
This direct communication allows better utilization of the dataflow characteristics of reconfigurable execution pipelines by directly integrating the communication into the pipelines.
One important tool to evaluate the performance characteristics of the different inter-\ac{FPGA} communication interfaces that are used in multi-\ac{FPGA} systems are benchmarks.

However, for the \ac{HPC} area, benchmark suites with relevant benchmark applications that allow the evaluation of these systems are rare.
To overcome this shortage, we earlier proposed \emph{HPCC FPGA}~\cite{hpcc_fpga} based on the HPC Challenge Benchmark suite \cite{HPCCIntroduction} targetting the HPC domain.
In this previous work, we focussed on the performance characterization of a single \ac{FPGA} with regards to memory access patterns of the applications.
Some important benchmarks of the HPC Challenge are missing in the proposed suite.
The missing benchmarks \emph{b\_eff} a synthetic network bandwidth benchmark, \emph{PTRANS} a parallel matrix transposition, and \emph{LINPACK} are good candidates to scale over multiple \acp{FPGA} and stress the communication interfaces.
For inter-\ac{FPGA} communication, there does not exist a standard comparable to \ac{MPI} on CPUs.
The serial interfaces of recent \acp{FPGA} can be used to establish circuit-switched and packet-switched networks.
This includes implementations of an Ethernet core \cite{FPGAEthernet} that can be directly used from \ac{HLS} code or application-specific protocols \cite{MLNetwork}.
Another approach is the Intel-specific \ac{OpenCL} extension for point-to-point connections \ac{IEC}.
With SMI \cite{SMI}, also a publicly available library for the communication in a \ac{CSN} based on \ac{IEC} has been proposed.
This approach abstracts away the routing but is still vendor-specific. %
In consequence, the only way of communication that is available for both -- Intel and Xilinx FPGAs -- and that is based on well-established standards is the data exchange via PCIe and MPI via the inter-CPU network.

Therefore, to create a widely usable benchmark suite for multi-FPGA systems, we make the following contributions:
\begin{itemize}
\item We extend the benchmark suite with three new communication-focused multi-\ac{FPGA} benchmarks, including LINPACK, and provide baseline implementations compatible with a wide range of FPGAs.
\item We add support for multi-FPGA execution and validation for all existing benchmarks of the suite and propose improved designs that allow a well-scaling execution over dozens of FPGAs.
\item We provide a vendor-specific, optimized implementation using \ac{IEC} for all three benchmarks to show the easy extendability of the benchmark suite with vendor-specific communication interfaces.
\item We evaluate all benchmarks on two different multi-\ac{FPGA} systems with Intel and Xilinx FPGAs. The results show that distributed FPGA systems can reach HPC performance and thus require corresponding benchmarking techniques.
\end{itemize}

We made the extensions of the suite publicly available and contributed the proposed changes to the official sources of the HPCC FPGA benchmark suite. \footnote[1]{\url{https://github.com/pc2/HPCC_FPGA}}

\section{Parallel Implementation of HPC Challenge Benchmarks for FPGA}
\label{sec:hpcc-benchmark}
The existing benchmark kernels of HPCC FPGA are called \emph{base implementations} and are designed to provide good performance on different FPGA architectures.
On the one hand, this is achieved with configuration parameters that allow to scale the benchmark kernels and, on the other hand, with code optimizations that apply to a broad range of FPGAs.
This allows the creation of efficient designs without manual code changes for different \ac{FPGA} architectures.
The configuration parameter \texttt{NUM\_REPLICATIONS} is supported by all benchmarks of the suite and is used to replicate kernels to increase resource utilization.
We also support this parameter in the newly added benchmarks.
A more detailed description of the build process is given in our previous work \cite{hpcc_fpga} and in the online documentation.~\footnote[2]{\url{https://pc2.github.io/HPCC_FPGA}}

Different hardware interfaces can be utilized for inter-\ac{FPGA} communication with recent \ac{FPGA} boards.
Direct inter-FPGA communication via the serial interfaces requires vendor-specific extensions and libraries, which makes it impossible to create \emph{base implementations} with this approach.
However, sending the data over the host via PCIe and MPI can be implemented in vendor-independent OpenCL code.
Thus, we use this communication approach in the base implementations.

\begin{figure}
    \centering
    \includegraphics[width=\linewidth]{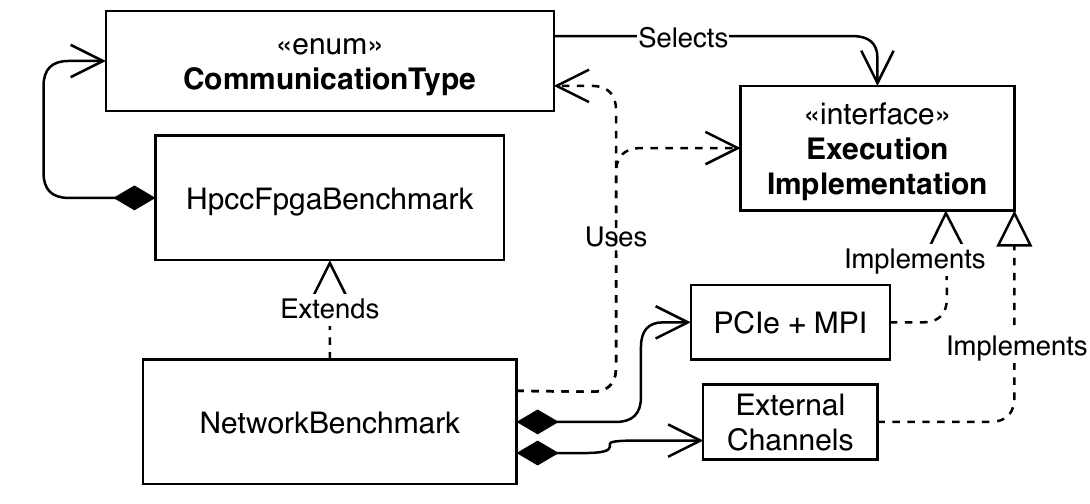}
    \caption{Improved architecture of the benchmark host code to increase extendability with different OpenCL kernels. \texttt{NetworkBenchmark} is the host implementation of one of the new benchmarks. It itself contains different implementations for the execution of the OpenCL kernels on the FPGA that depend on the communication scheme. To extend a benchmark for another communication scheme, only a new execution implementation needs to be added.}
    \label{fig:hpcc_fpga_execution_types}
\end{figure}

Since we use the same structure for code organization and the build process as the existing benchmarks in \emph{HPCC FPGA}, the new benchmarks come with support for custom kernels.
This allows easy extension of the benchmarks with additional OpenCL kernels.
One restriction is that the OpenCL kernels need to have the same kernel signature to work with the existing host code.
For some communication schemes it may also be required to slightly change the kernel signature to pass additional data to the kernels.
We extended the host code architecture as shown in Figure~\ref{fig:hpcc_fpga_execution_types} to also simplify the extension of the benchmarks from the host side.
The \texttt{CommunicationType} defines the different communication schemes that are supported by the benchmark suite.
Each benchmark inherits from the \texttt{HpccFpgaBenchmark} class, so a substantial part of the host code is shared between all benchmark host implementations.
The actual OpenCL kernel execution is done in the implementations of the \texttt{ExecutionImplementation} interface.
This enables FPGA designs with different number of OpenCL kernels or kernel signatures.
During runtime, the execution implementation is selected based on the \texttt{CommunicationType} which is determined by the name of the bitstream file.
Adding support for another FPGA design with different kernels only requires an additional implementation of the execution interface as it is done for \emph{PCIe + MPI} and \emph{External Channels} in the class diagram.
In this work, we use this feature to implement optimized versions of the benchmarks for Intel FPGAs with direct inter-FPGA communication via QSFP ports.
The OpenCL extension that we use for this implementation is called \ac{IEC}.

For every benchmark, there has to be one \ac{MPI} rank per \ac{FPGA}, so the number of \ac{MPI} ranks needs to match the number of used \acp{FPGA}.
Before every kernel execution on the FPGA, the hosts synchronize using an MPI barrier to reduce the measurement error.
Always the slowest execution time among all FPGAs is reported for each repetition of the benchmark execution.
The best repetition is used to calculate the derived performance metric of the benchmark.

In the following, we give a detailed description of the new benchmarks implementations.

\subsection{Effective Bandwidth Benchmark}

In this benchmark, we use the rules of the Effective Bandwidth (b\_eff) benchmark given in \cite{b_eff_website}.
It is a synthetic benchmark that uses the derived metric \emph{effective bandwidth} to combine both -- the network latency and bandwidth -- into a single metric.
The original benchmark sends messages of sizes $2^0, 2^1, \dots, 2^{20}$ \si{\byte} to neighbor nodes in a ring topology.
The effective bandwidth is calculated from the measured bandwidth for the different message sizes as shown in Equation~\ref{eq:effective_bandwidth}.

\begin{equation}
    b_{\mathit{eff}} = \frac{\sum_L(\mathit{max}_{\mathit{rep}}(b(L, \mathit{rep})))}{21}
    \label{eq:effective_bandwidth}
\end{equation}

where $L$ are the message sizes, $\mathit{rep}$ the repetitions of the execution and $b(L, \mathit{rep})$ the measured bandwith for message size $L$ during repetition $\mathit{rep}$.

The base implementation of this benchmark does not require a \ac{FPGA} kernel because data is transferred between the \acp{FPGA} solely by the host.
The optimized version for Intel FPGAs is configurable with the parameters given in Table~\ref{tab:b_eff_configuration_parameters}.
Next to the number of kernel replications, it contains the width of the external channels in Bytes.

\begin{table}
    \centering
    \caption{Configuration parameters of the b\_eff benchmark}
    \begin{tabular}{p{2.5cm}p{4.5cm}}
    \toprule
    \textbf{Parameter} & \textbf{Description} \\
    \midrule
\texttt{CHANNEL\_WIDTH} & The width of a single external channel in bytes \\
         \bottomrule
    \end{tabular}
    \label{tab:b_eff_configuration_parameters}
\end{table}

\subsubsection{Base Implementation}

The base implementation exchanges the messages between the global memory of neighboring \acp{FPGA} in the ring.
Therfore, it reads a memory buffer representing the message using the OpenCL directive \texttt{clEnqueueReadBuffer} from \ac{FPGA} to host.
In a second step, it exchanges the buffer via \texttt{MPI\_Sendrecv} with the node that contains the neighboring \ac{FPGA}.
In a last step, the buffers are written to the global memory of the \acp{FPGA} with the \texttt{clEnqueueWriteBuffer} OpenCL directive.
These steps are executed for both directions in the ring and for all message sizes.

The expected performance is limited by the required time to read ($\mathit{pcie\_read}_t$) and to write ($\mathit{pcie\_write}_t$) a message of the given size to the FPGA via PCIe, plus the time required to exchange the message between the nodes using MPI ($\mathit{mpi}_t$).
All three steps need to be executed sequentially, so the expected bandwidth for a message size $L$ can be modelled with Equation~\ref{eq:b_eff_exe_time_pcie}.

\begin{equation}
    b_{L} = \frac{2 \cdot L}{\mathit{pcie\_write}_t + \mathit{mpi}_t + \mathit{pcie\_read}_t}
    \label{eq:b_eff_exe_time_pcie}
\end{equation}

\subsubsection{Intel External Channels Implementation}
\label{sec:intel_external_channels_implementation}

The Intel-optimized implementation requires OpenCL kernel code and consists of two different kernel types: a \emph{send} kernel and a \emph{receive} kernel.
During execution, they continuously send or receive data over two external channels of the width specified in \texttt{CHANNEL\_WIDTH}.
A kernel replication always consists of both kernels since a \emph{send} kernel always requires a counterpart.
Because the message size might exceed the width of the channels, the messages are further divided into data chunks that match the channel width.
Thus, a message is streamed chunk-wise over two channels to the receiver in a pipelined loop.
At kernel start, the \emph{send} kernel will generate a message chunk that is filled with bytes of the value $ld(m) \mod 256$.
The message chunk will be used continuously for sending and will be stored in global memory after the last transmission.
This allows verifying the correct transmission of the data chunk over the whole range of repetitions.

\begin{figure}
    \centering
    \includegraphics[width=\linewidth]{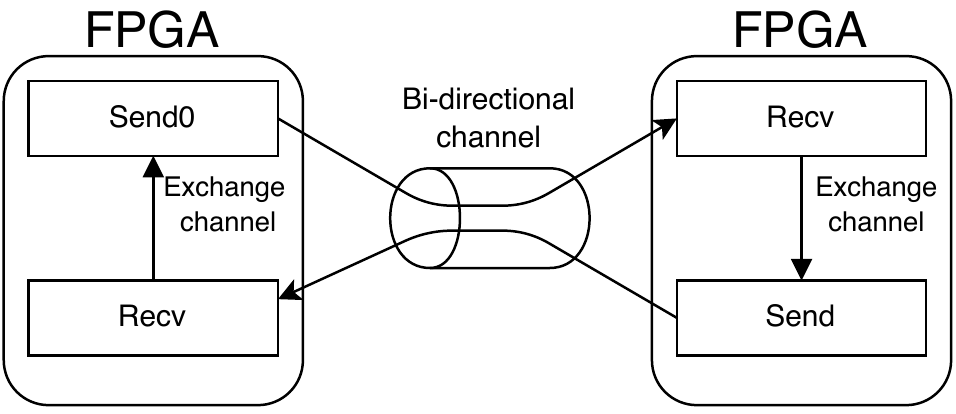}
    \caption{Data exchange of two kernel pairs over the external channels. The kernel pairs are executed on two different FPGAs and communicate over the bi-directional external channels. The kernel pairs are connected over a internal channel to forward the received data to the send kernel for the next iteration.}
    \label{fig:b_eff_kernels}
\end{figure}

A schematic view of the channel connections for the kernel implementation is given in Figure~\ref{fig:b_eff_kernels}.
The kernels are connected to another kernel pair on a different \ac{FPGA}.
In the Figure, the kernels form a small ring over two FPGAs and the topology can be arbitrarily scaled by adding more FPGAs.

The arrows describe the path of a single data chunk through the kernels.
A message chunk will be repeatedly sent over the external channel until the sum of all sent chunks matches the desired message size.
Messages are sent in parallel in both directions.
After a complete message is sent, the message chunk is forwarded from the receive to the send kernel over the internal channel.
Only then, the next message is sent now using the message chunk received over the internal channel.
The message chunk is stored in a global memory buffer after the last message is exchanged and used for validation on the host side.
A single send-receive kernel pair will use two external channels in both directions.

\begin{table}
    \centering
    \caption{Characteristics of the serial channel IP of the BittWare 520N board taken from the specification \cite{bittware_bsp_doc}}
    \begin{tabular}{p{1.5cm}p{3.5cm}>{\raggedleft\arraybackslash}p{1cm}@{\hskip 1pt}p{1cm}}
    \toprule
    \textbf{Parameter} & \textbf{Description} & \multicolumn{2}{c}{\textbf{Value}} \\
    \midrule
$c_n$ & Number of external channels & 4 &\\
$c_l$ & Latency of a channel & 520 & \si{\nano\second} \\
$c_f$ & Frequency of a channel & 156.25 & \si{\mega\hertz} \\
$c_w$ & Width of a channel & 32 & \si{\byte} \\
         \bottomrule
    \end{tabular}
    \label{tab:520n_external_channels}
\end{table}

The performance metric of the benchmark combines latency and the total bandwidth of the network.
To model the performance, we need precise information about the latency and bandwidth of the external channels as they are given in Table~\ref{tab:520n_external_channels}.
Every kernel replication can only make use of two external channels, which means that $c_n' = 2$ and the total number of external channels is utilized by using a replication count of 2.
The execution time of a kernel pair for a given message size can then be modeled with Equation~\ref{eq:b_eff_exe_time} where $L$ is the used message size and $i$ the number of messages that are sent.

\begin{equation}
    t_{L, i} = \frac{\lceil\frac{L}{c_n' \cdot c_w}\rceil \cdot i}{c_f} + i \cdot c_l
    \label{eq:b_eff_exe_time}
\end{equation}

For the bandwidth model, we insert the values for the IP core taken from Table~\ref{tab:520n_external_channels} which results in Equation~\ref{eq:b_eff_bandwidth}.

\begin{equation}
    b_{L} = \frac{2 \cdot L}{\lceil\frac{L}{64B}\rceil \cdot 6.4ns + 520ns}
    \label{eq:b_eff_bandwidth}
\end{equation}

This equation models the bandwidth for a single send-receive kernel pair and is expected to scale linearly with the number of kernel pairs.

\subsection{Parallel Matrix Transposition}

\begin{figure}
    \centering
    \includegraphics[width=\linewidth]{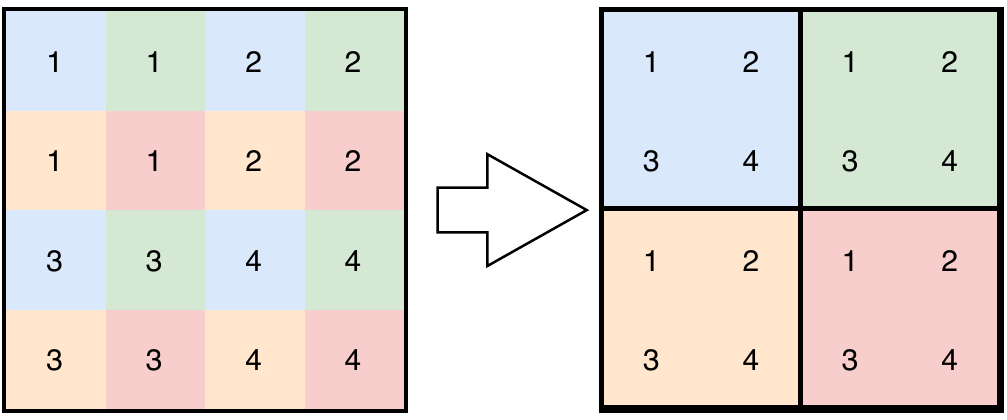}
    \caption{Diagonal distribution of the 16 blocks of a $4 \times 4$ block matrix on four FPGAs with $P=Q=2$. The original matrix is shown on the left. Colors equal the FPGA in which global memory the data block of the matrix will reside at the beginning of the calculation. On the right, the placement of the data on the different FPGAs is shown. The bold lines represent the borders of the memory of a single FPGA.}
    \label{fig:ptrans_data_distribution}
\end{figure}

The parallel matrix transposition (PTRANS) benchmark computes the solution of $C = B + A^T$ where $A,B,C \in \mathbb R^{n \times n}$.
The matrix $A$ is transposed and added to another matrix $B$. The result is stored in matrix $C$.
All matrices are divided into blocks, and the blocks are distributed over multiple \acp{FPGA} using a PQ distribution scheme shown in Figure~\ref{fig:ptrans_data_distribution}.

\begin{table}
    \centering
    \caption{Configuration parameters of the PTRANS benchmark}
    \begin{tabular}{p{2.5cm}p{4.5cm}}
    \toprule
    \textbf{Parameter} & \textbf{Description} \\
    \midrule
\texttt{BLOCK\_SIZE} & Size of the matrix blocks that are buffered in local memory and also distributed between the FPGAs. \\
\rule{0pt}{3ex}%
\texttt{CHANNEL\_WIDTH} & Width of the channels in data items. Together with the used data type, the width of the channel in bytes can be calculated. \\
\rule{0pt}{3ex}%
\texttt{DATA\_TYPE} & Specifies the used data type for the calculation. \\
         \bottomrule
    \end{tabular}
    \label{tab:ptrans_configuration_parameters}
\end{table}

\subsubsection{Base Implementation}

The configuration parameters for the implementation are given in Table~\ref{tab:ptrans_configuration_parameters}.
The number of kernel pairs can be defined with the \texttt{NUM\_REPLICATIONS} parameter and the block size that also defines the sizes of the local memory buffers can be set with \texttt{BLOCK\_SIZE}.
Moreover, the width of the channel is defined by \texttt{CHANNEL\_WIDTH} and \texttt{DATA\_TYPE}.
It should match the width of the used communication channel.
In case of the base implementation, this is the width of the global memory interface.

The implementation consists of a single OpenCL kernel that sequentially executes three pipelines for every matrix block.
In the first pipeline, a block of matrix $A$ is read from global memory and written into a buffer.
The second pipeline reads the block of $A$ transposed from the buffer, reads a block of $B$ from global memory, adds both blocks, and stores the result in an additional buffer.
The content of this buffer is written back to global memory in the last pipeline.

Every pipeline is reading or writing a single block of data from the global memory.
This is a similar approach as it is used for the STREAM benchmark in the suite and leads to an efficient use of the global memory.
Before the kernel can be executed, the matrix $A$ needs to be exchanged by the host ranks using \texttt{MPI\_Sendrecv}.

With Equation~\ref{eq:ptrans_performance}, the expected execution time for a single matrix block is given.
It consists of the time required to exchange the blocks via MPI and write them into the global memory of the \acp{FPGA} ($t_\mathit{MPI}$) and the execution time of the OpenCL kernel.
The kernel execution time is based on the three pipelines that are executed sequentially, the block size $b$, the channel width in number of values $c_w$, and the clock frequency of the used channel $c_f$.
Depending on the number of kernel replications, the \ac{FPGA} may be able to process multiple matrix blocks simultaneously without interference.

\begin{equation}
    t_{\mathit{PTRANS}} = t_\mathit{MPI} + 3 \cdot \frac{b^2}{c_w} \cdot c_f
    \label{eq:ptrans_performance}
\end{equation}

For the verification of the data, the non-transposed blocks of matrix $A$ are exchanged by the hosts using MPI.
Then, each host re-calculates the result using a CPU reference implementation.
The reported error is the maximum residual error between the FPGA and CPU result.

\subsubsection{Intel External Channels Implementation}

The Intel-specific implementation comes with the restriction that $P\overset{!}{=}Q$.
This allows to set up a static circuit-switched network between the pairs or \acp{FPGA} and exchange the matrix blocks without additional routing.
The FPGA logic is implemented in two kernels per external channel, similar to the b\_eff benchmark.
For this implementation, the width of the channel defined by \texttt{CHANNEL\_WIDTH} and \texttt{DATA\_TYPE} should match the width of the external channels.

One of the kernels reads a block of $A$ into local memory. The size of this local memory buffer can be defined with \texttt{BLOCK\_SIZE}.
The block of matrix $A$ is then read transposed from local memory and written into the external channel.
Reading from global memory and writing to the external channel is implemented in a single pipeline, double buffering the local memory block.

The second kernel will receive chunks of a transposed block of $A$, add a block of $B$ to it and store it in global memory.
In consequence, no local memory is needed in this kernel.
One major goal of this implementation is to continuously send and receive data over all available external channels to utilize the available network bandwidth, which is most likely the performance bottleneck.
Nevertheless, the kernels may also suffer from low global memory bandwidth because they need to concurrently read and write to three different buffers for every kernel replication.

This leads to a total required global memory bandwidth on a single \ac{FPGA} given in Equation~\ref{eq:ptrans_mem_bandwidth}.
\begin{equation}
    b_{\mathit{global}} = 3 \cdot r \cdot c_w \cdot c_f
    \label{eq:ptrans_mem_bandwidth}
\end{equation}
where $r$ the number of external channels per FPGA (or number of kernel replications), and $c_f$ and $c_w$ the frequency and the width of an external channel as defined in Table~\ref{tab:520n_external_channels}.
This means the required global memory bandwidth is three times higher than the network bandwidth to keep the benchmark network-bandwidth-bound.
As a performance metric, the \ac{FLOP} per second are calculated.
For the calculation it is assumed, that $n^2$ additions are required for the computation on matrices of width $n$.
Considering the characteristics of the external channels of the used BittWare 520N boards, the maximum performance will be $p = i \cdot r \cdot 32B \cdot 156.25MHz$ for an sufficiently large matrix, where $i$ is the number of used \acp{FPGA}.
Note, that the block size is not considered in this performance model.
It is used to allow larger memory bursts from global memory, which are defined by the width of the block.
This will lead to a higher efficiency of the global memory accesses, but since the performance model covers the case where the network bandwidth is the bottleneck, this parameter can be neglected.
However, for very small block sizes, the efficiency of the global memory may be reduced to the point that it becomes the bottleneck.

\subsection{High Performance Linpack}

The High Performance Linpack benchmark solves a large equation system $A \cdot x = b$ for $x$, where $A \in \mathbb R^{n \times n}$ and $b,x \in \mathbb R^{n}$.
This is done in two steps: First the matrix $A$ is decomposed into a lower matrix $L$ and an upper matix $U$.
In a second step, these matrices are used to first solve $L \cdot y = b$ and finally $U \cdot x = y$ to get the result for the vector $x$.
For the implementation of the benchmark on \ac{FPGA}, the rule set for the HPL-AI mixed-precision benchmark \cite{hpl_ai} was adapted, which defines $A$ to be a diagonally dominant matrix.
Thus, the LU factorization does not require pivoting.
In contrast to the original benchmark it is possible to choose between single-precision and double-precision floating-point values.
Since the benchmark suite is designed to only measure the \ac{FPGA} performance, no additional iterative method is used to refine the result if a lower precision is used.
Only the LU decomposition, which is the most compute-intensive step in this calculation, is executed on the \acp{FPGA}.
The number of \acp{FLOP} for this step is defined to be $\frac{2 \cdot n^3}{3}$ for a matrix $A$ with width $n$ in contrast to $2 \cdot n^2$ for solving the equation systems for the LU-decomposed matrix.
Only the performance of the LU factorization on the FPGA is reported.

\begin{figure}
    \centering
    \includegraphics[width=0.7\linewidth]{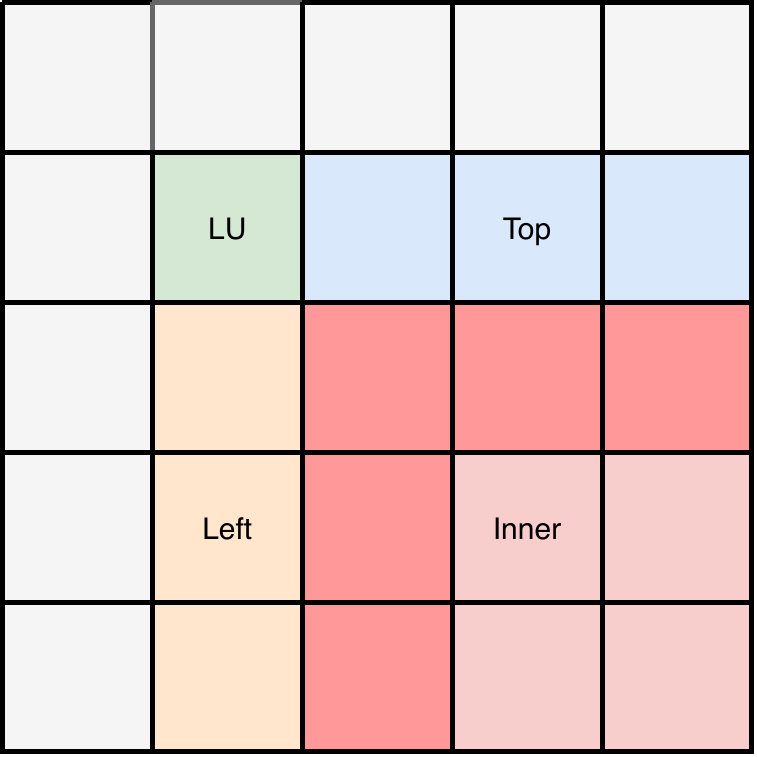}
    \caption{In every iteration of the algorithm, a single block in the matrix is decomposed into a lower and upper matrix (green). The lower matrix is used to update all blocks on the right of this block (blue) and the upper matrix to update all blocks below this block (orange). The updated Top and Left blocks are then used to update all inner blocks (red) and the dark red blocks need to be updated before the next communication phase can start.}
    \label{fig:lu_iteration}
\end{figure}

\subsubsection{Base Implementation}

The base implementation uses a blocked, right-looking variant for the LU factorization as it is described in \cite{blocked_lu_factorization}.
Therefore, the matrix will be divided into sub-blocks with a width of $2^\mathit{BLOCK\_SIZE\_LOG}$ elements.
The exact size of the blocks is defined over a configuration parameter.
For the update of a single row and column of blocks, we need to perform four different operations.
A single iteration of the blocked LU decomposition is shown in Figure~\ref{fig:lu_iteration}.
In every iteration, the LU factorization for a diagonal block of the matrix is calculated which is marked green in the visualization.
All grey-colored blocks on the left and top of this block are already updated in previous iterations and will require no further processing.
This is why this approach is called right-looking, since we will always update the blocks on the right of the LU block.
After the LU block is decomposed, the lower matrix block $L$ is used to update all blocks on the right of the LU block.
Since they are the top-most blocks that still require an update, they are in the following called \emph{top blocks}.
The upper matrix block $U$ is used to update all blocks below the current LU-block.
These are the left-most blocks that require an update, so they are referred to as \emph{left blocks}.
The left and top blocks again are used to update all inner blocks, which can efficiently be done using matrix multiplication.
The design contains a separate kernel for each of the four operations.

Additionally, a single iteration of the LU decomposition is split into two subsequent steps in the design.
In the \emph{communication phase}, the LU, left, and top blocks are updated which also involves data exchange between kernels on the same \ac{FPGA} and between the \acp{FPGA}.
In the \emph{update phase}, the exchanged data is used to update all inner blocks locally using matrix multiplication kernels.

\begin{figure}
    \centering
    \includegraphics[width=\linewidth]{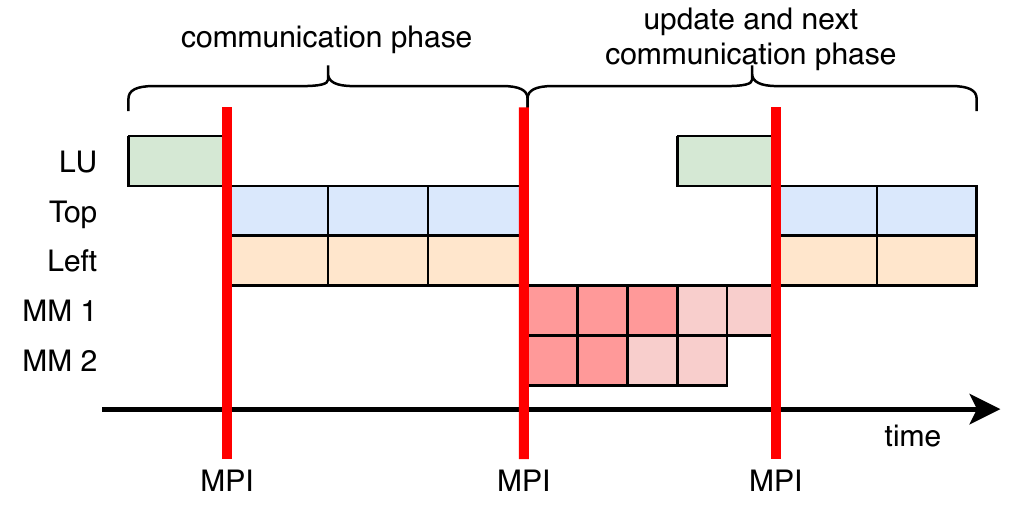}
    \caption{Kernel executions over time for a single iteration of the LU decomposition in the base implementation. During the communication phase, data needs to be exchanged two times between FPGAs using MPI and PCIe. The matrix multiplication kernels are executed in the update phase. Communication and update phase of subsequent iterations can overlap, so communication latency can partially be hidden.}
    \label{fig:lu_execution_schedule_pcie}
\end{figure}

Both phases can overlap as shown by the timeline of kernel executions in Figure~\ref{fig:lu_execution_schedule_pcie} based on the matrix given in Figure~\ref{fig:lu_iteration}.
The number of matrix multiplications required for a single iteration of the algorithm increases quadratically with the matrix size.
No data dependency exists between the light red matrix multiplications and the operations of the next \emph{communication phase}, which allows overlapping of the two phases.
For large matrices this means that the performance of the implementation is limited by the aggregated performance of the matrix multiplication kernels.
During the \emph{communication phase}, matrix blocks are exchanged via the host using PCIe and MPI.

\subsubsection{Intel External Channels Implementation}

\begin{figure}
    \centering
    \includegraphics[width=\linewidth]{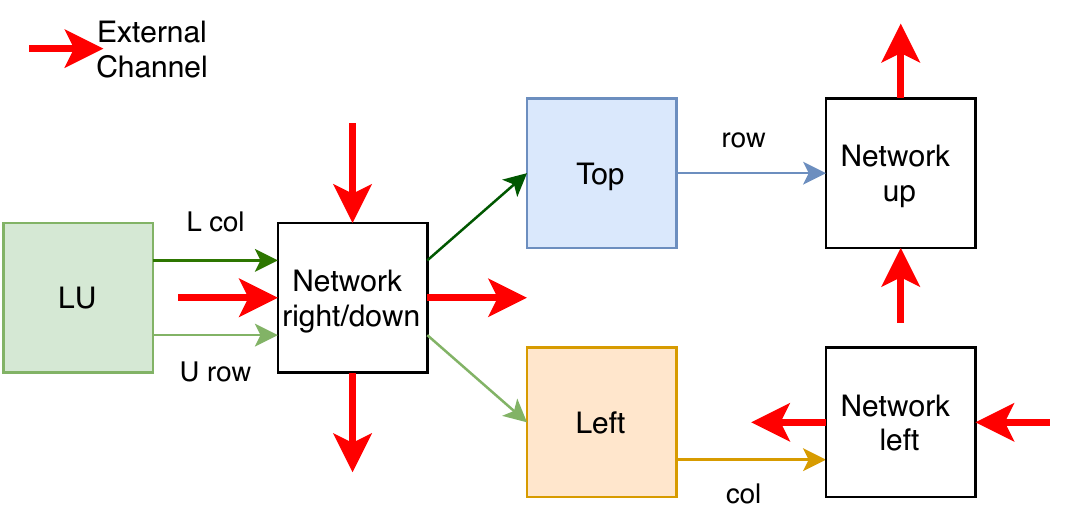}
    \caption{Data flow through the kernels of the communication phase on a single FPGA. The kernels are connected over internal channels. Between the calculation kernels, network kernels are used to select the correct input for the next kernel from the internal or external channels. The network kernels for the top and left direction will also store incoming matrix blocks as input for the matrix multiplication. The bold arrows represent the serial channels and their direction in the 2D torus.}
    \label{fig:lu_communication_kernels}
\end{figure}

Figure~\ref{fig:lu_communication_kernels} shows connections between the kernels used in the communication phase.
For the execution over multiple FPGAs, the boards are arranged into a quadratic 2D torus of variable size using the point-to-point connections.
Not all kernels need to be active on every \ac{FPGA} within a single iteration.
Instead, data can also be received over the external channels if it is computed on another \ac{FPGA}.
If the FPGA is in charge of calculating the LU block, the LU kernel is executed and the decomposed L and U blocks are forwarded row and column-wise to a network kernel.
The network kernel forwards the data over the external channels to neighboring \acp{FPGA} in the torus.
The four possible directions are used for different types of data as it is indicated by the red arrows.
Moreover, the network kernel forwards the locally computed L and U block to the Left and Top kernel.
The top and left kernel use the data to update a block with the L or U block and forward the updated block to the next network kernels.
Here, the input data is selected either from the internal or external channels and data is forwarded over the external channels if required.
Besides that, incoming data is stored in global memory buffers for later use in the update phase.
By splitting the network communication into three kernels, it is possible to establish a cycle-free data path through the torus during the communication phase.
This reduces the impact of pipeline and channel latencies during this phase.

\begin{figure}
    \centering
    \includegraphics[width=\linewidth]{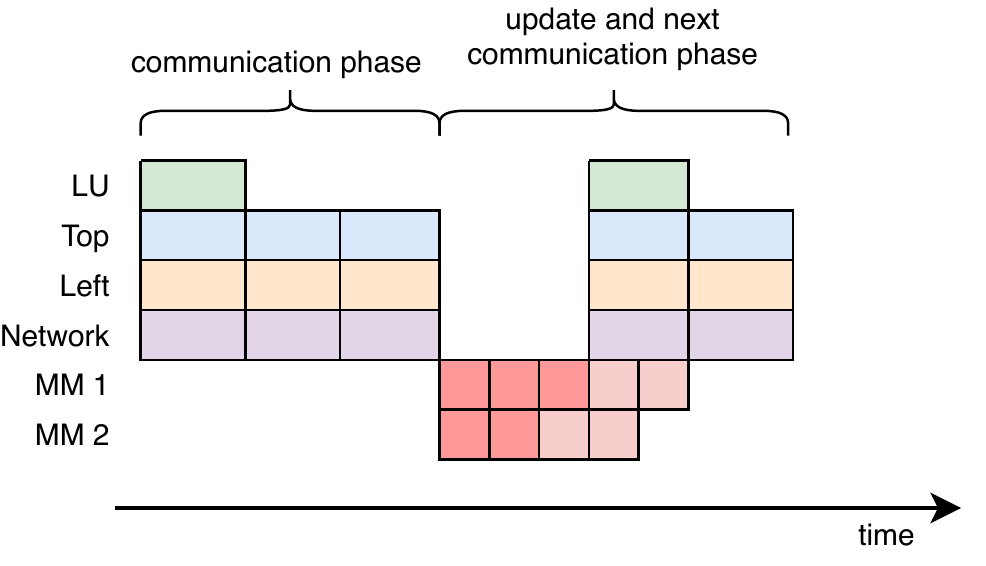}
    \caption{Kernel executions over time for a single iteration of the LU decomposition. During the communication phase, the network kernels are active whereas during the update phase the matrix multiplication kernel is executed. Communication and update phase of subsequent iterations overlap.}
    \label{fig:lu_execution_schedule}
\end{figure}

During the transfer from the left kernel to the network kernel, the left blocks are transposed.
This allows a simplified design of the matrix multiplication for the inner blocks since all input matrices can be processed row-wise.
Figure~\ref{fig:lu_execution_schedule} shows a part of the execution of the kernels over time for the iteration given in Figure~\ref{fig:lu_iteration}.
It can be seen that the LU kernel is only executed once per communication phase.
The lower and upper matrices are buffered by the left and top kernel to allow the update of subsequent blocks.
All network kernels are summarized under \emph{Network} in the graph.
In the example, two matrix multiplication kernels are used, and the blocks a redistributed between the two replications.
The next communication phase starts as soon as the first row and column of the inner blocks is updated, which is represented by the dark red blocks.

\begin{figure}
    \centering
    \includegraphics[width=\linewidth]{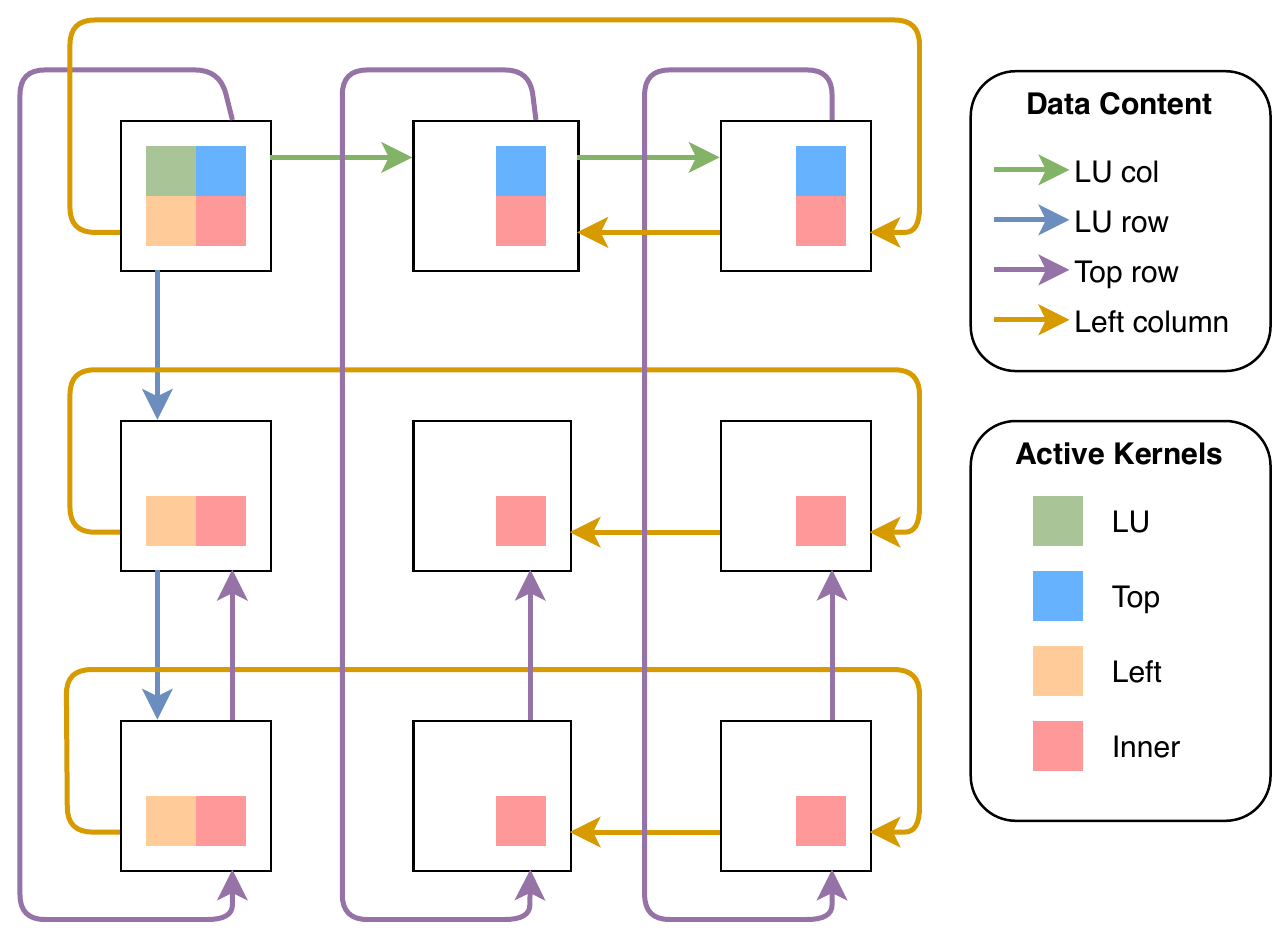}
    \caption{For the LU decomposition, the FPGAs use a 2D torus network topology to exchange data. This example shows the active kernels during a single iteration in a $3 \times 3$ torus. The black boxes are the FPGAs, and the colors within the boxes indicate the active kernels. The direction and type of data that is forwarded between the FPGAs is given by the arrows. \rev{In every iteration of the algorithm this communication scheme shifts one FPGA to the bottom-right in the torus.}}
    \label{fig:lu_torus_data_exchange}
\end{figure}

A 2D torus is used to connect multiple \acp{FPGA} for the LU decomposition.
Every \ac{FPGA} is programmed with the same bitstream and the host schedules the kernels in the required order and configuration.
Matrix blocks are distributed between the \acp{FPGA} using a PQ-grid of the size of the torus.
This allows balancing the load between the devices more evenly since the matrix will get smaller with every iteration of the algorithm.
In Figure~\ref{fig:lu_torus_data_exchange} the active kernels and the data exchange between \acp{FPGA} in a $3 \times 3$ torus is shown for a global matrix size of more than 12 blocks so \ac{FPGA} has to update more than four blocks.
In this case, only the \ac{FPGA} on the top left needs to execute all four compute kernels, but every FPGA will use its matrix multiplication kernel.

The base implementation of the HPL benchmark uses a similar two-leveled blocked approach than the GEMM benchmark described in \cite{hpcc_fpga}.
Thus, it uses two parameters to specify the block sizes of the local memory buffers and of the compute units as described in Table~\ref{tab:hpl_configuration_parameters}.
Additionally, it is possible to specify the data type and specify the number of matrix multiplication kernels using the \texttt{NUM\_REPLICATIONS} parameter.
The two-leveled blocked approach is also used for the communication kernels and all kernels use the same first level block sizes.
A main difference is the parallelism in the computation between communication kernels and matrix multiplication.
The matrix multiplication kernels unroll the computation in three dimensions in the second level whereas the communication only use a two-dimensional unrolling.
This means the parallel calculation increase cubically with the chosen register block size for the matrix multiplications and only quadratically for the communication kernels.
It would also be possible to scale the second level blocks differently between the two kernel types.
This has to be considered in further optimizations steps.

\begin{table}
    \centering
    \caption{Configuration parameters of the LINPACK benchmark}
    \begin{tabular}{p{3cm}p{4.5cm}}
    \toprule
    \textbf{Parameter} & \textbf{Description} \\
    \midrule
\texttt{LOCAL\_MEM\_BLOCK\_LOG} & Logarithm of the size of the matrix blocks that are buffered in local memory and also distributed between the FPGAs. \\
\rule{0pt}{3ex}%
\texttt{REGISTER\_BLOCK\_LOG} & Logarithm of the size of the second level matrix blocks. The kernels contain completely unrolled logic to start the computation of such a sub-block every clock cycle. \\
\rule{0pt}{3ex}%
\texttt{DATA\_TYPE} & Specifies the used data type for the calculation. \\
         \bottomrule
    \end{tabular}
    \label{tab:hpl_configuration_parameters}
\end{table}

Only the LU factorization of matrix $A$ is calculated on the FPGAs.
After this step, the equation system is solved using a distributed CPU reference implementation among all MPI ranks.
The input matrix was generated such that the resulting vector is a vector of all ones.
The reported error is the normalized maximum residual error calculated with $\frac{||x||}{n \cdot ||b|| \cdot \epsilon}$ where $n$ is the width of matrix $A$ and $\epsilon$ the machine epsilon.

\subsection{Extend Existing Benchmarks for Multi-FPGA Execution}

In addition to the new benchmarks proposed in this paper, we also extend the existing benchmarks of our previous work \cite{hpcc_fpga} for the execution in a multi-\ac{FPGA} environment.
An essential configuration parameter for all benchmarks is the specification of kernel replications \texttt{NUM\_REPLICATIONS}.
These kernel replications are kernels with the same or very similar functionality to allow a higher utilization of the \ac{FPGA} resources.
The input data is then distributed between the replicated kernels such that every kernel works on its own data set.
Especially for devices with \ac{HBM2} this step is crucial to make use of the high number of memory banks and the high aggregated bandwidth.
So the existing implementations in HPCC FPGA already handle the memory banks on a single \ac{FPGA} like a distributed memory system.
As a result, changes on the \ac{OpenCL} kernels are not required for most of the legacy benchmarks, and the extensions focus on the host codes to trigger the distributed execution and support the validation and result collection over multiple compute nodes using \ac{MPI}.

The RandomAccess benchmark was not well scalable because it could at best update a single data item per clock cycle even when scaled over multiple \acp{FPGA}.
This is limited by the way the pseudo-random numbers for the address calculation are generated.
We now allow the generation of multiple pseudo-random numbers per clock cycle to overcome this limitation by replicating the \ac{RNG}.
This also changed the configuration parameters of the benchmark as given in Table~\ref{tab:ra_configuration_parameters}.

\begin{figure}
    \centering
    \includegraphics[width=\linewidth]{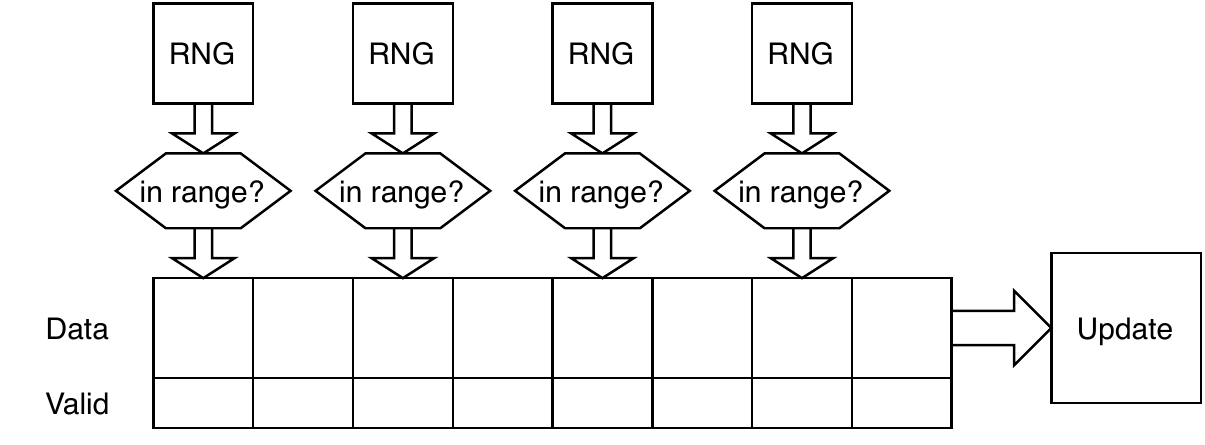}
    \caption{RandomAccess shift register used to connect the different \acp{RNG}.}
    \label{fig:ra_shift_register}
\end{figure}

A single replication of the improved RandomAccess kernel is given in Figure~\ref{fig:ra_shift_register}.
The \acp{RNG} are initialized with different seeds to generate a sub-part of the random-number sequence.
In consequence, the same random number as with the old version are generated, only the order of updates may vary.
Every clock cylce, the \ac{RNG} outputs a new random number.
This number is placed into a shift register, if two conditions hold:

\begin{enumerate}
    \item The buffer address derived from the random number is in range of the kernel replication.
    \item The shift register does not already contain a valid random number at the position where it should be inserted.
\end{enumerate}

In the latter case, the \ac{RNG} will stall until the random number can be placed into the shift register.
So in other terms, the produced random numbers are sequentialzed by the shift register for the input into the actual update logic.
This approach increases the probability, that the update logic processes a valid address for high numbers of kernel replications.
Since scaling over multiple \acp{FPGA} corresponds to increasing the number of kernel replications, this does also improve the performance in multi-\ac{FPGA} execution.

\begin{table}
    \centering
    \caption{Configuration parameters of the RandomAccess benchmark}
    \begin{tabular}{p{4cm}p{3.5cm}}
    \toprule
    \textbf{Parameter} & \textbf{Description} \\
    \midrule
    \texttt{HPCC\_FPGA\_RA\_DEVICE\_ BUFFER\_SIZE\_LOG} & Logarithm of the size of the data buffer that is randomly updated in number of values.\\
    \rule{0pt}{3ex}%
    \texttt{HPCC\_FPGA\_RA\_RNG\_COUNT\_LOG} & Logarithm of the number of \acp{RNG} that are created per kernel replication.\\
    \rule{0pt}{3ex}%
    \texttt{HPCC\_FPGA\_RA\_RNG\_DISTANCE} & Distance between \acp{RNG} in the shift register.\\
    \bottomrule
    \end{tabular}
    \label{tab:ra_configuration_parameters}
\end{table}

\section{Benchmark Execution and Evaluation}
\label{sec:evaluation}
In the following, we execute and evaluate the scaling behavior of the existing and the three new benchmarks of the benchmark suite on two different multi-\ac{FPGA} systems containing Xilinx or Intel \acp{FPGA}.

\subsection{Evaluation Setup and Synthesis Results}

For the evaluation of the benchmarks we used version 0.5.1 of the benchmark suite \cite{hpcc_fpga_release} and we made all artifacts and code modifications for additional experiments publicly available \cite{hpcc_fpga_evaluation}.

We synthesized and executed the benchmarks on two multi-\ac{FPGA} systems: The Noctua system of \ac{PC$^2$} at Paderborn University and the Xilinx FPGA evaluation system of the Systems Group at ETH Zurich.
The Noctua system uses \emph{Nallatech/BittWare 520N} boards equipped with Intel Stratix 10 GX2800 \acp{FPGA}, where every node is a two-socket system equipped with Intel Xeon Gold 6148 CPUs, \SI{192}{\giga\byte} of DDR4-2666 main memory, and two \acp{FPGA} connected via x8 PCIe 3.0.
Moreover, the nodes in the cluster communicate over a hybrid network:
The CPUs use an Intel Omni Path network with \SI{100}{\giga\bit\per\second} per port, whereas the \acp{FPGA} can exchange data over the four serial interfaces with up to \SI{40}{\giga\bit\per\second} per port.
The serial interfaces are connected to a CALIENT S320 Optical Circuit Switch that allows the configuration of arbitrary full-duplex point-to-point connections between the serial interfaces of the \acp{FPGA}.
This functionality allows creating the desired network topology and is used in the following to establish connections between up to 26 \acp{FPGA} to execute the optimized versions of the benchmarks.
The network topology is set up before running the benchmarks and stays unchanged during execution.

\ac{BSP} version 20.4.0 and Intel OpenCL \ac{SDK} for FPGA version 21.2.0 are used to synthesize all benchmark kernels.
This \ac{BSP} version comes in two different sub-versions with and without support for the external channels.
All benchmarks were synthesized with the \emph{HPC} sub-version which offers no support for communication over the external channels.
This \ac{BSP} requires slightly fewer resources than the \emph{MAX} sub-version with external channel support.
Only the optimized versions of b\_eff, PTRANS, and LINPACK are synthesized with the \emph{MAX} sub-version.
The host codes are compiled with GCC 8.3.0 and Intel MPI 2019 Update 6 Build 20191024. Configuration and generation of the build scripts are done using CMake 3.15.3.

Additionally, the benchmarks are synthesized and executed on \emph{Xilinx Alveo U280} boards.
The Xilinx FPGA evaluation system contains four of these boards.
As \ac{SDK}, Vitis 2020.2 is used with the shell \emph{xilinx\_u280\_xdma\_201920\_3} and XRT 2.9.
The host codes are compiled with GCC 7.5.0 and OpenMPI 2.1.1.
Each FPGA is controlled by an Intel(R) Xeon(R) Gold 6234 CPU with \SI{108}{\giga\byte} of main memory.

All benchmark kernels are designed to be independent of the number of used \acp{FPGA}, so only a single synthesis for every benchmark and FPGA board is required for the evaluation.
With configuration parameters, it is possible to improve resource utilization and performance of the benchmark kernels for a specific \ac{FPGA} board before synthesis.
For the benchmarks STREAM, FFT, and GEMM, these configuration parameters were discussed in more detail in \cite{hpcc_fpga}.
Table~\ref{tab:used_parameters} contains the used configurations for each benchmark.
The configuration parameters are chosen to better utilize performance-relevant resources on the \ac{FPGA}.

In Table~\ref{tab:resource_usage} the resource usage of the synthesized benchmark kernels is given.
Our updated scalable implementation of the RandomAccess benchmark now requires additional logic and BRAMs to implement the \acp{RNG}.
We have chosen the number of \acp{RNG} in the configuration to be the next larger power of two of the used \acp{FPGA}.
This increaes the probability, that a valid number can be processed in every clock cycle by every FPGA.
Further increasing the number of \acp{RNG} may lead to lower clock frequencies, wich will also have an impact on performance.
With the recent SDK version we were able to synthesize the GEMM benchmark with a much higher clock frequency for the BittWare 520N, which promises a large performance improvement.
For the b\_eff benchmark, only a single bitstream is synthesized.
For the base implementations, no bitstream is required since the data transfer is solely handled by the host.
Also resource consumption is not an issue with this synthetic benchmark because the performance is mainly limited by the network bandwidth and latency.

PTRANS requires BRAM buffers that are used to block-wise transpose the matrix and store intermediate results. A large block size will benefit large memory bursts but it is also important to achieve a clock frequency close to 300~MHz to make best use of the memory bandwith.
For LINPACK it is -- similar to GEMM -- important to maximize the number of used DSPs for matrix multiplications. In addition, mainly some extra BRAM is required to store the matrix blocks for the kernels of the communication phase.
As a result, the resource utilization is very similar to GEMM. A significant difference is visible for the Alveo U280, where only 69\% of the DSPs can be utilized.
We were not able to fit the communication kernels and three matrix multiplication kernels on the FPGA, so we had to reduce the number of matrix multiplication kernels. Since we are not able to scale the matrix multiplication kernels individually, this lead to a onsiderable amount of unused resources.

The difference in DSPs between the base and optimized implementation for the BittWare 520N is caused by the way, the multiplication of the $8 \times 8$ matrices in registers is implemented.
In case of the \ac{IEC} version, only multiply-adds are used consuming 512 DSPs in total per replication.
In the base implementation, the compiler created the same matrix multiplication from 64 dot-products of size 8 followed by 64 additions. This slightly increases the DSP usage to 576 DSPs per replication but considerably reduces the logic and BRAM usage as it can be seen in the resource utilization.

Besides the two bitstreams for the baseline and one for the vendor-specific implementation, we also synthesized LINPACK with a block size of 256 elements for the 520N.
This is the same block size that is used on the U280 and allows a better comparison of the performance results of both \ac{FPGA} boards.
The configuration requires considerably lesser logic and BRAM compared to the version wih 512 element block width and achieves a higher clock frequency.

\begin{table*}
    \centering
    \caption{Synthesis configurations of all benchmarks}
    \begin{tabular}{p{2cm}p{5.5cm}>{\raggedleft\arraybackslash}p{1cm}>{\raggedleft\arraybackslash}p{1cm}>{\raggedleft\arraybackslash}p{1cm}>{\raggedleft\arraybackslash}p{1cm}}
    \toprule
        \textbf{Benchmark} &
        \textbf{Parameter} &
        \textbf{520N IEC} &
        \textbf{520N PCIe} &
        \textbf{U280 PCIe}\\
         \midrule
  \multirow{5}{*}{STREAM} &\texttt{NUM\_REPLICATIONS} & 4 & 4 & 2 \\
  &\texttt{DATA\_TYPE} & float & float & float\\
  &\texttt{GLOBAL\_MEM\_UNROLL}  & 1 & 1& 1\\
&\texttt{VECTOR\_COUNT} & 16 & 16 & 16\\
&\texttt{DEVICE\_BUFFER\_SIZE} & 32,768 & 32,768 & 16,384\\
\midrule
\multirow{4}{*}{RandomAccess}  & \texttt{NUM\_REPLICATIONS} & 4 & 4 & 2\\
&\texttt{HPCC\_FPGA\_RA\_DEVICE\_BUFFER\_SIZE\_LOG} & 0 & 0 & 10\\
&\texttt{HPCC\_FPGA\_RA\_RNG\_COUNT\_LOG} & 5 & 5 & 3\\
&\texttt{HPCC\_FPGA\_RA\_RNG\_DISTANCE} & 5 & 5 & 1\\
 \midrule
\multirow{2}{*}{FFT} &\texttt{NUM\_REPLICATIONS} & 2 & 2 & 1\\
& \texttt{LOG\_FFT\_SIZE}  & 17 & 17 & 9 \\   
\midrule
  \multirow{5}{*}{GEMM} &\texttt{NUM\_REPLICATIONS} & 5 & 5 & 3 \\
  &\texttt{DATA\_TYPE} & float & float & float\\
  &\texttt{GLOBAL\_MEM\_UNROLL}  & 8 & 8& 8 \\
  &\texttt{BLOCK\_SIZE} & 512 & 512 & 256 \\
&\texttt{GEMM\_SIZE} &  8 & 8 & 8\\
\midrule
  \multirow{2}{*}{b\_eff} &\texttt{NUM\_REPLICATIONS} & 2 & \multicolumn{2}{c}{only host code required} \\
  &\texttt{CHANNEL\_WIDTH}  & 8 & \multicolumn{2}{c}{only host code required} \\
\midrule
  \multirow{4}{*}{PTRANS} &\texttt{NUM\_REPLICATIONS} & 4 & 4 & 2\\
  &\texttt{DATA\_TYPE} & float & float & float\\
  &\texttt{CHANNEL\_WIDTH}  & 8 & 16 & 16\\
  &\texttt{BLOCK\_SIZE} & 512 & 512 & 256 \\
\midrule
  \multirow{4}{*}{LINPACK} &\texttt{NUM\_REPLICATIONS} & 5 & 5 & 2\\
  &\texttt{DATA\_TYPE} & float & float & float\\
  &\texttt{LOCAL\_MEM\_BLOCK\_LOG}  & 9 & 9 & 8\\
  &\texttt{REGISTER\_BLOCK\_LOG} & 3 & 3 & 3\\
         \bottomrule
    \end{tabular}
    \label{tab:used_parameters}
\end{table*}

\begin{table*}
    \centering
    \caption{Resource usage of the synthesized benchmarks}
    \begin{tabular}{p{2cm}p{1cm}>{\raggedleft\arraybackslash}p{1cm}@{\hskip 3pt}p{1cm}>{\raggedleft\arraybackslash}p{1cm}@{\hskip 3pt}p{1cm}>{\raggedleft\arraybackslash}p{1cm}@{\hskip 3pt}p{1cm}p{1cm}>{\raggedleft\arraybackslash}p{1cm}}
    \toprule
       \textbf{Benchmark} &
       \textbf{Device} &
       \multicolumn{2}{c}{\textbf{Logic}} &
       \multicolumn{2}{c}{\textbf{BRAM}} &
       \multicolumn{2}{c}{\textbf{DSPs}} &
       \textbf{Freq. [MHz]} & \textbf{Comp. Time [h]} \\
    \midrule
        \multirow{2}{*}{STREAM} & 520N & 178,268 & (19\%) & 3,926 & (33\%) & 128 & (2\%) & 341.67 & 2.64\\
                             & U280 & 188,124  & (14\%) & 854 & (42\%) & 170 & (2\%) & 300.00 & 2.44\\
                             \midrule
        \multirow{2}{*}{RandomAccess} & 520N & 222,405 & (24\%) & 602 & (5\%) & 14 & (<1\%) & 325.00 & 3.50\\
                                  & U280 & 184,888 & (14\%) & 350 & (17\%) & 24 & (<1\%) & 300.00 & 2.20\\
                                  \midrule
        \multirow{2}{*}{FFT} & 520N & 280,105 & (30\%) & 1,811 & (15\%) & 1,560 & (27\%) & 400.00 & 4.10\\
                          &U280 & 375,069  & (29\%) & 342 & (17\%) & 682 & (8\%) & 286.00 & 6.94\\
                          \midrule
        \multirow{2}{*}{GEMM} & 520N & 310,564 & (33\%) & 8,321 & (71\%) & 3,318 & (58\%) & 272.50 & 9.87\\
                            & U280 & 659,200 & (51\%) & 1,139 & (57\%) & 7,714 & (85\%) & 186.00 & 12.25\\
                            \midrule
        \multirow{3}{*}{b\_eff} & 520N$^2$ & 173,010& (19\%) & 512 & (4\%) &  0& (0\%) & 290.63 & 2.45\\
               & 520N$^1$ & \multicolumn{8}{c}{only host code required}\\
               & U280$^1$ & \multicolumn{8}{c}{only host code required}\\
        \midrule
        \multirow{3}{*}{PTRANS} & 520N$^2$ & 242,232 & (26\%) & 4,756 & (41\%) & 68 & (1\%) & 281.25 & 3.64\\
                            & 520N$^1$ & 233,317 & (25\%) & 4,662 & (40\%) & 162 & (3\%) & 380.00 & 3.62\\
                            & U280$^1$ & 283,028  & (22\%) & 598 & (30\%) & 96 & (1\%) & 283.00 & 4.07\\
                            \midrule
        \multirow{3}{*}{LINPACK} & 520N$^2$& 361,377 & (39\%) & 8,326 & (71\%) & 2,809 & (49\%) &  225.00 & 10.40\\
                              & 520N$^1$& 303,471 & (33\%) & 8,245 & (70\%) & 3,185 & (55\%) &  233.34 & 9.42\\
                              & 520N$^{1,3}$& 276,546 & (30\%) & 2,587 & (22\%) & 3,185 & (55\%) &  280.00 & 6.08\\
                              & U280$^1$& 663,418 & (51\%) & 994 & (49\%) & 6,201 & (69\%) &  156.00 & 12.10\\
    \bottomrule
    \multicolumn{10}{r}{$^1$communication via MPI and PCIe using the host network}\\
    \multicolumn{10}{r}{$^2$communication via Intel external channels (IEC) OpenCL extension}\\
    \multicolumn{10}{r}{$^3$version with a reduced block size of 256 elements}\\
    \end{tabular}
    \label{tab:resource_usage}
\end{table*}

\subsection{Evaluation of the Effective Bandwidth and PTRANS}

The \emph{b\_eff} benchmark does not only report the derived metric \emph{effective bandwidth} but also the achieved bandwidth for all tested message sizes, which range from a single byte to 1~MB.
A new message is only sent after the current message is received by the neighbor node.

The base implementation of the b\_eff benchmark reads the data from the \ac{FPGA} board to the host using an OpenCL call, exchange the data between the host CPUs using the \texttt{MPI\_Sendrecv} method, and write the data back to the \ac{FPGA} using OpenCL.
This means, in contrast to the optimized \ac{IEC} implementation, no OpenCL device code is required for this benchmark to work.

\begin{figure}
  \includegraphics[width=\linewidth]{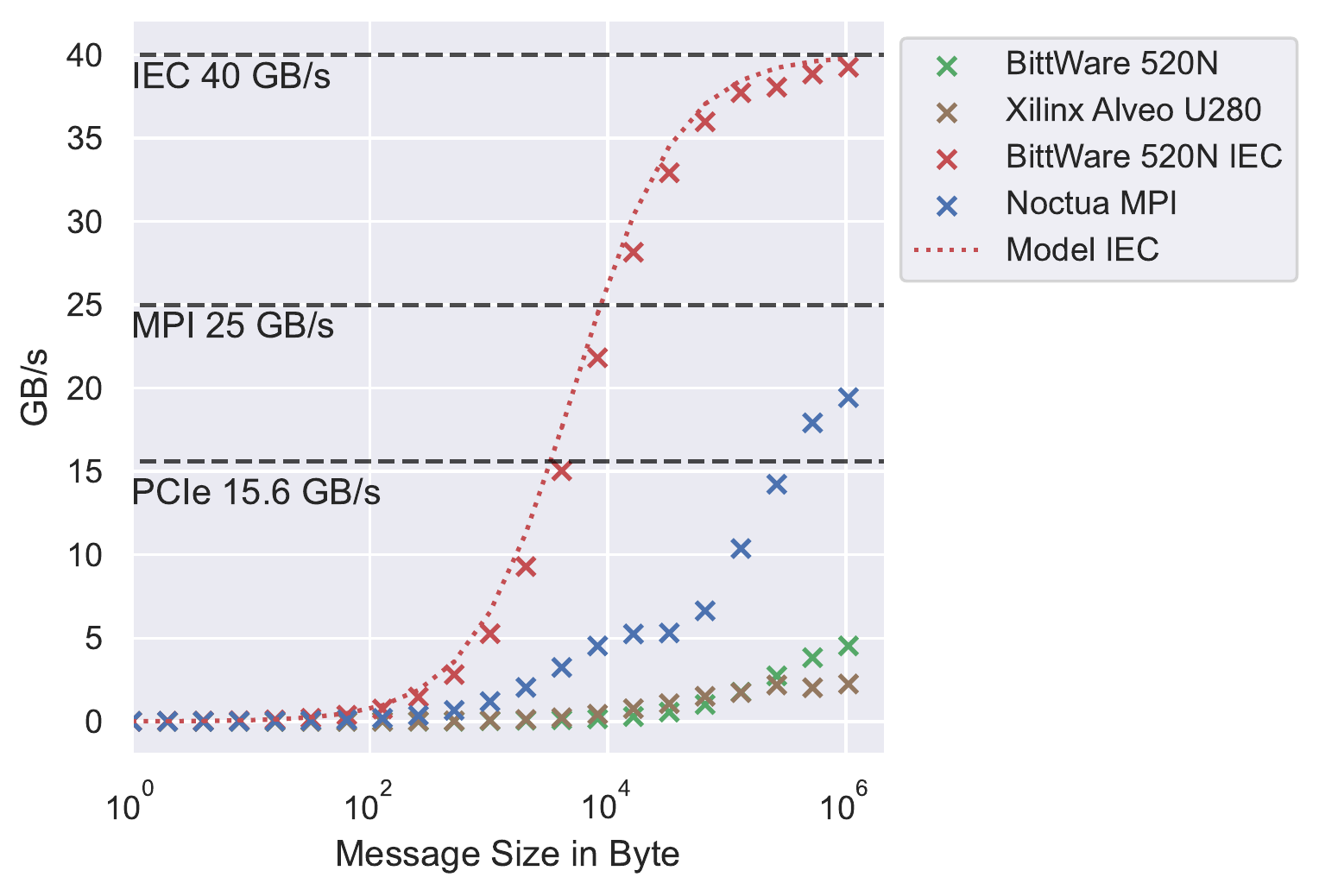}
  \caption{The aggregated bandwidth over different message sizes measured by the b\_eff benchmark over two CPUs or FPGAs. Next to the measurements, the maximum performance for the communication between FPGAs, CPUs, and between FPGAs and CPUs via PCIe.}
  \label{fig:beff_message_sizes_mpi}
\end{figure}

The measured total bandwidth over the message sizes is given in Figure~\ref{fig:beff_message_sizes_mpi} for two \acp{FPGA} or CPU nodes respectively.
We do not show the MPI-only performance for the Xilinx system in this plot since it heavily overlaps with the measurements for the base implementations.
The maximum theoretical bandwidths for PCIe, MPI via the Intel Omni Path 100Gbps interconnect, and \ac{IEC} are given by the black dashed lines in the plot.
For the base implementations, the bandwidth remains below \SI{5}{\giga\byte\per\second} on both devices although the theoretical bandwidth of PCIe and MPI are both much higher.
The bandwidth gets limited by the additional copy operations that are required to get the data from FPGA to CPU and back.
This leads to a theoretic peak performance of \SI{5.9}{\giga\byte\per\second} for this approach.
However, the measurements with the MPI-only implementation of the benchmark show, that the maximum message size of \SI{1}{\mega\byte} is not sufficiently large to utilize the MPI peak performance on Noctua.

The optimized \ac{IEC} approach shows maximum bandwidths close to the theoretical peak for \SI{1}{\mega\byte} message sizes.
Also the measurements closely correlate to the model described in Section~\ref{sec:intel_external_channels_implementation}.

\begin{figure}
  \includegraphics[width=\linewidth]{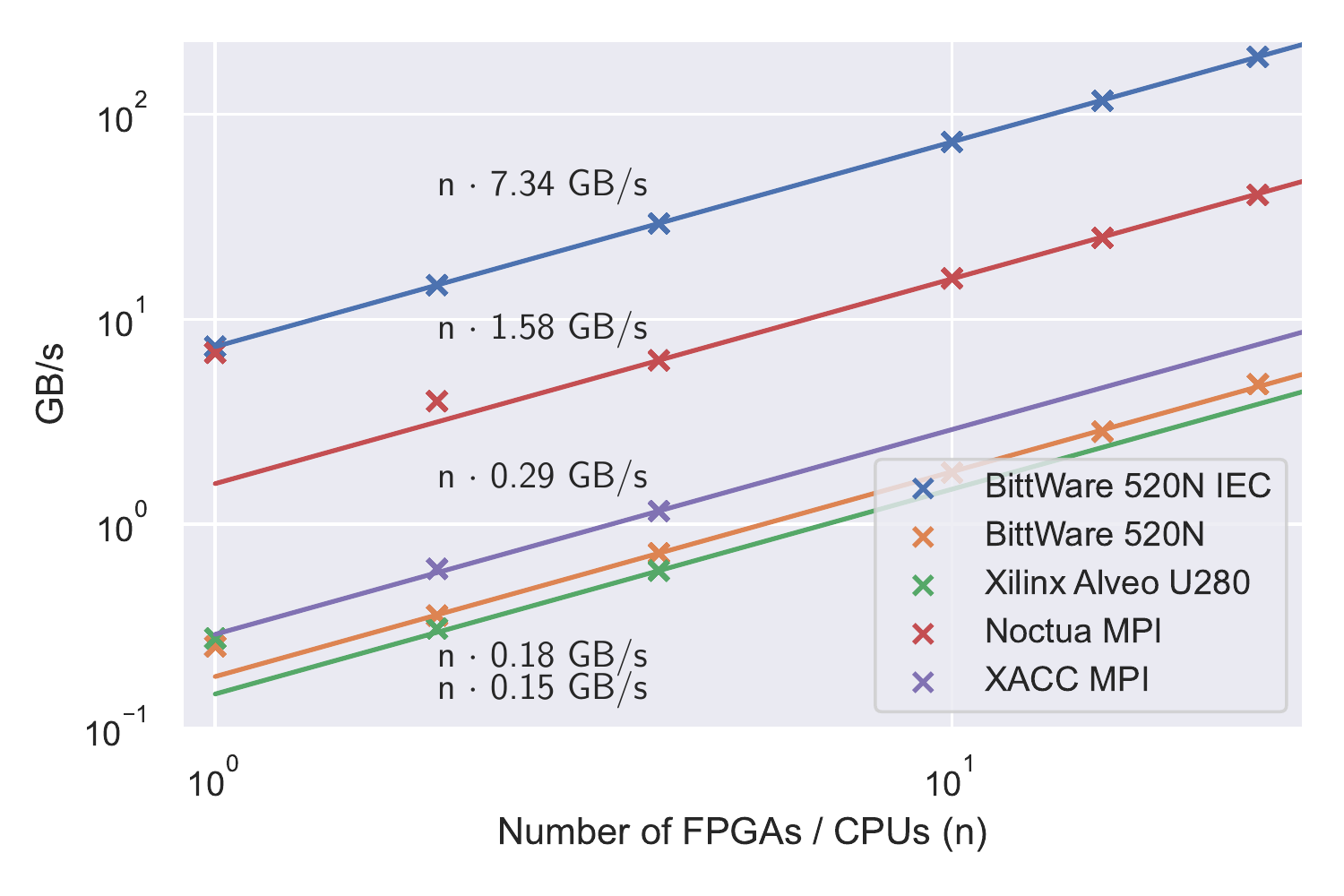}
  \caption{The measured effective bandwidth over the number of used FPGAs and CPUs. A logarithmic scale is used for the ring size and the measured bandwidth. The colored lines represent the perfect scaling based on measured effective bandwidth over two FPGAs or CPU nodes.}
  \label{fig:beff_effective_bw}
\end{figure}

The linear scaling behavior for the derived \emph{effective bandwidth} metric is visualized for all for implementations in Figure~\ref{fig:beff_effective_bw}.
All implementations show nearly perfect scaling with the available network bandwidth as it is indicated by the extrapolation lines for each device.
Especially for the MPI-only and the MPI + PCIe version, the performance on a single node is considerably higher than the available network bandwidth because in these cases the data will be transferred between the ranks using shared memory.
We also observe a huge difference in the effective bandwidth between the two MPI versions of Noctua and the Xilinx system.
These measurements have to be kept in mind when comparing the results of the base implementation on the two systems, since the huge difference in MPI performance will also have an impact on the PCIe + MPI performance.
The ability to add additional optimized implementations of the benchmark kernels allows to generate comparable results not only limited to \acp{FPGA} but also for CPU or other accelerators.
This only requires minor changes in the existing code base and large amounts of the code can be reused including handling of input parameters, input data generation and validation, calculation of derived metrics, and printing of performance summaries.

\begin{figure*}
  \includegraphics[width=0.65\linewidth]{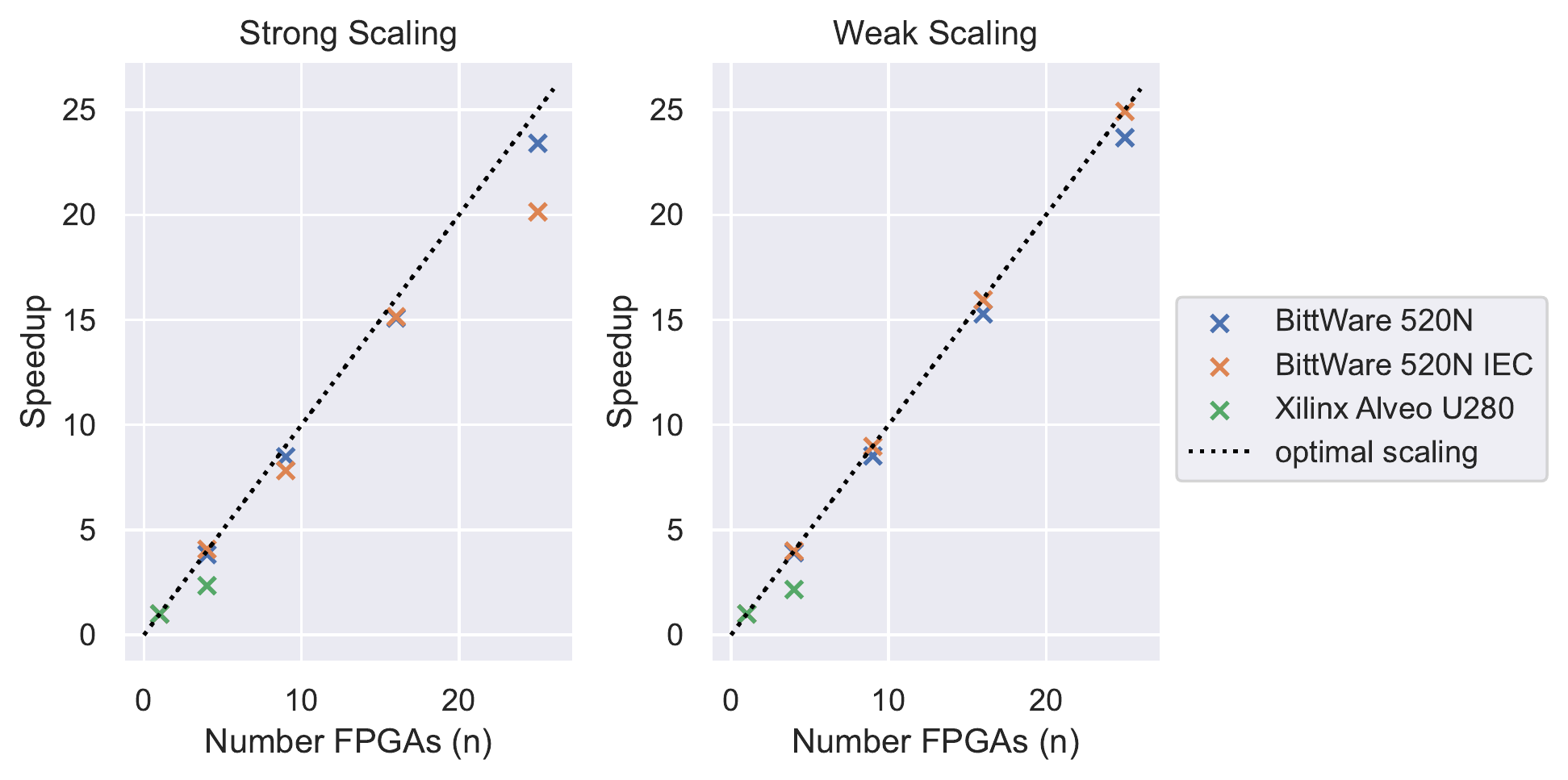}
  \caption{Speedup of the PTRANS benchmark executed with a quadratic matrix of 16,384 elements over up to 25 FPGAs in a weak and strong scaling scenario.}
  \label{fig:ptrans_performance}
\end{figure*}

For the matrix transposition, the blocks of a matrix are distributed among the \acp{FPGA} in a PQ distribution where $P = Q$.
A matrix of 32,768 elements is transposed using strong and weak scaling. 
The resulting speedups for the different \acp{FPGA} is given in Figure~\ref{fig:ptrans_performance}.
In the strong scaling experiment, the base implementation on the BittWare 520N shows a better scaling behavior as the optimized version using \ac{IEC}.
This is because the base implementation is mainly bottlenecked by the PCIe bandwidth for the exchange of the matrices.
In contrast to that, the optimized version shows a significant reduction of the speedup for larger number of \acp{FPGA}.
This is caused by the compute pipeline on the \acp{FPGA} which can not be fully utilized with the smaller matrix sizes.

In the weak scaling experiment, the matrix size per \ac{FPGA} stays the same and the implementation achieves optimal speedup for up to 25 \acp{FPGA}.
The base implementation shows no significant differences for both \acp{FPGA} in strong and weak scaling.
On the Xilinx Alveo U280, the base implementation does not scale well. 
This is related to the low \ac{FPGA}-to-\ac{FPGA} bandwidth that we also measured in the b\_eff benchmark.
This means that this difference is caused by the comparably low MPI performance on the Xilinx system and not by the \acp{FPGA}.

\subsection{Evaluation of HPL}

\begin{figure}
  \includegraphics[width=\linewidth]{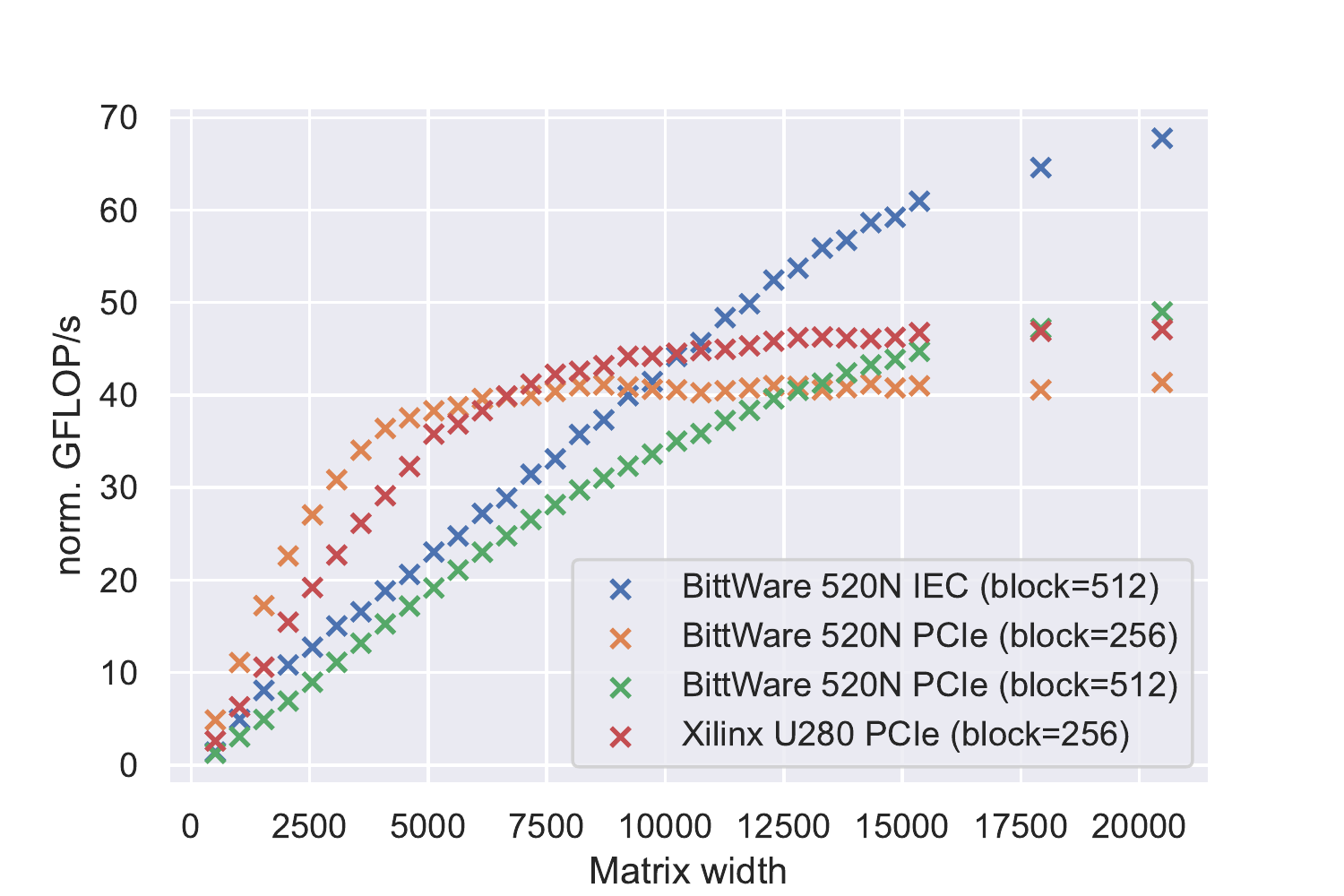}
  \caption{Normalized performance of the HPL bitstreams on the target FPGAs for different matrix sizes. The base versions of the benchmark are marked with PCIe, referring to the path of communication.  For small matrices, the design is limited by the communication latency until it can be efficiently hidden by matrix multiplications. Moreover, an additional execution for the BittWare 520N with a block size of 256 is given for comparison with the Xilinx Alveo U280.}
  \label{fig:hpl_scaling_single_fpga}
\end{figure}

In a first experiment, we measure the performance on a single \ac{FPGA} for different matrix sizes of up to 20,480 elements.
The performance for four bitstreams on the two different systems are given in Figure~\ref{fig:hpl_scaling_single_fpga}.
To allow an easier comparison of the efficiency of the design on the different platforms, the performance was normalized to a kernel frequency of \SI{100}{\mega\hertz} and a single kernel replication.
So the nomalized performance for a given matrix size should be very similar on the different platforms.
For small matrix sizes, the communication phase can not be overlapped with the computation phase.
Only for larger sizes of the matrix, both phases overlap for most of the computation time and the performance converges to the matrix multiplication performance.
Still, significant differences between the different bitstreams can be observed.
They are mainly caused by the chosen benchmark configuration parameters and compiler flags.
When comparing the base version and the vendor-specific version with Intel external channels (IEC) used on the BittWare 520N board, the base version using PCIe for communication shows lower performance, although the same configuration parameters are used.
The base version of the benchmark failed to synthesize with memory interleaving because additional \acp{LSU} are used for the communication and increase the complexity of the memory system.
Since only a single buffer is used to store matrix $A$, this effectively reduces the global memory bandwidth and stalls of the matrix multiplication kernels increase. 

On the Xilinx Alveo U280 board the largest block size that fits on the device is 256 elements in contrast to 512 element blocks for the 520N board.
The reduced block size results in an overlap of communication and computation for smaller matrix sizes but also reduces the peak performance because the utilization of the matrix multiplication pipeline decreases.
For comparison, we synthesized a bitstream for the 520N with a block size of 256.
It shows a similar scaling  behavior with regards to the matrix size but shows a slightly lower normalized peak performance.
The bitstream for the 520N achieves a nearly 80~\% higher frequency which also increases the memory bandwidth utilization and leads to more frequent pipeline stalls, which eventually leads to a lower nomalized performance.

\begin{figure}
  \includegraphics[width=\linewidth]{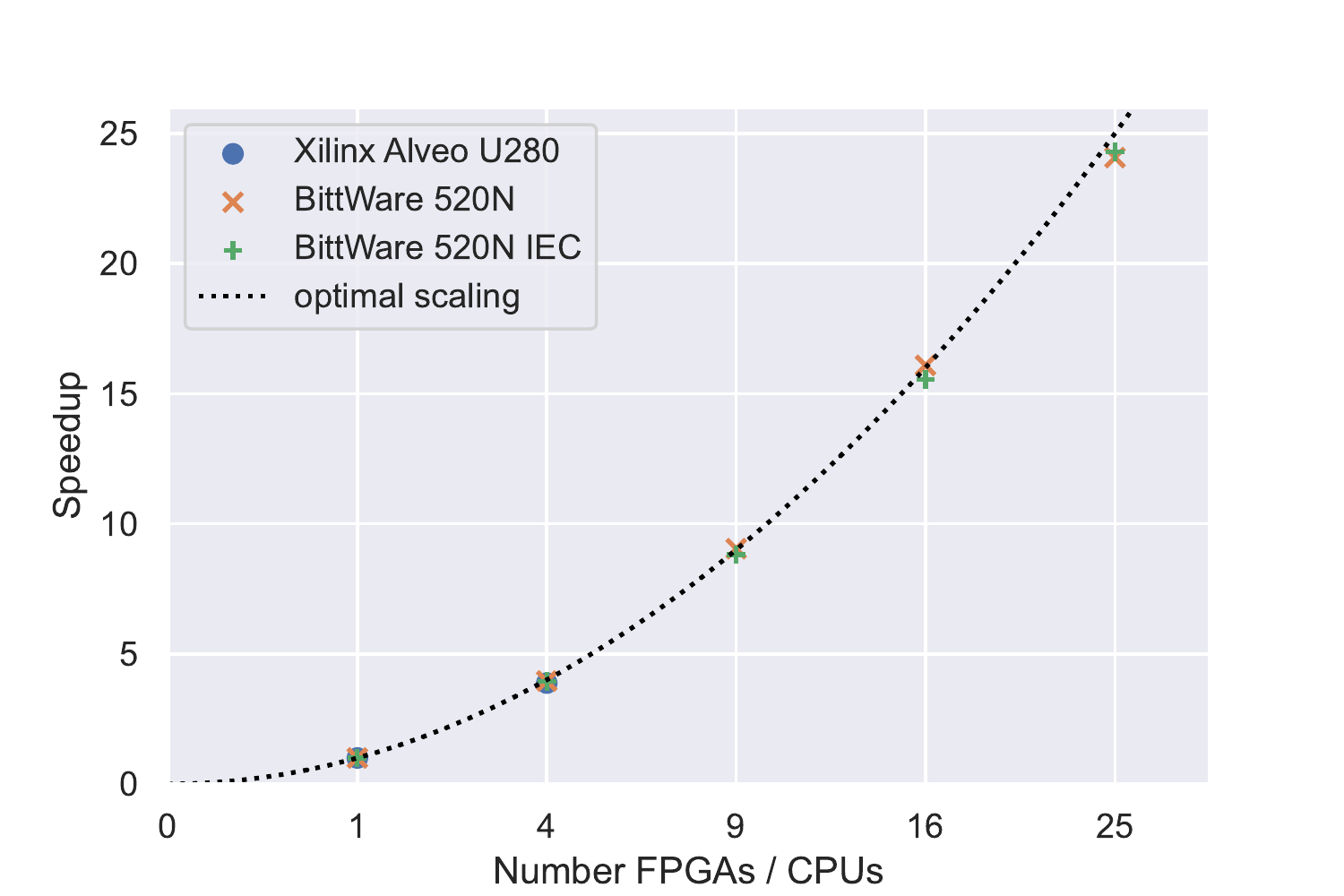}
  \caption{Speedup of HPL with a matrix width of 24,576 elements over multiple FPGAs in a weak scaling scenario.}
  \label{fig:hpl_torus_speedup_weak}
\end{figure}

Based on the measurements done with a single \ac{FPGA}, we set the  matrix size for the multi-\ac{FPGA} experiments to 24,576 elements, since all bitstreams will be close the their peak performance for this size.
We use this matrix size as a base for a weak scaling experiment, where the matrix size increases with the width of the \ac{FPGA} torus so the matrix size on a single \ac{FPGA} remains constant.
Additionally, we execute a strong scaling experiment, where the global matrix size remains constant while increasing the torus size.
The measurement results for the weak scaling experiment are given in Figue~\ref{fig:hpl_torus_speedup_weak}.
It can be observed, that all three implementations of the benchmark show a close to optimal scaling for up to 25 \acp{FPGA}.
Considering the differences in the network bandwidth that had an effect on the PTRANS results, this also means, that the benchmark implementation is compute bound on all \acp{FPGA}.

\begin{figure}
  \includegraphics[width=\linewidth]{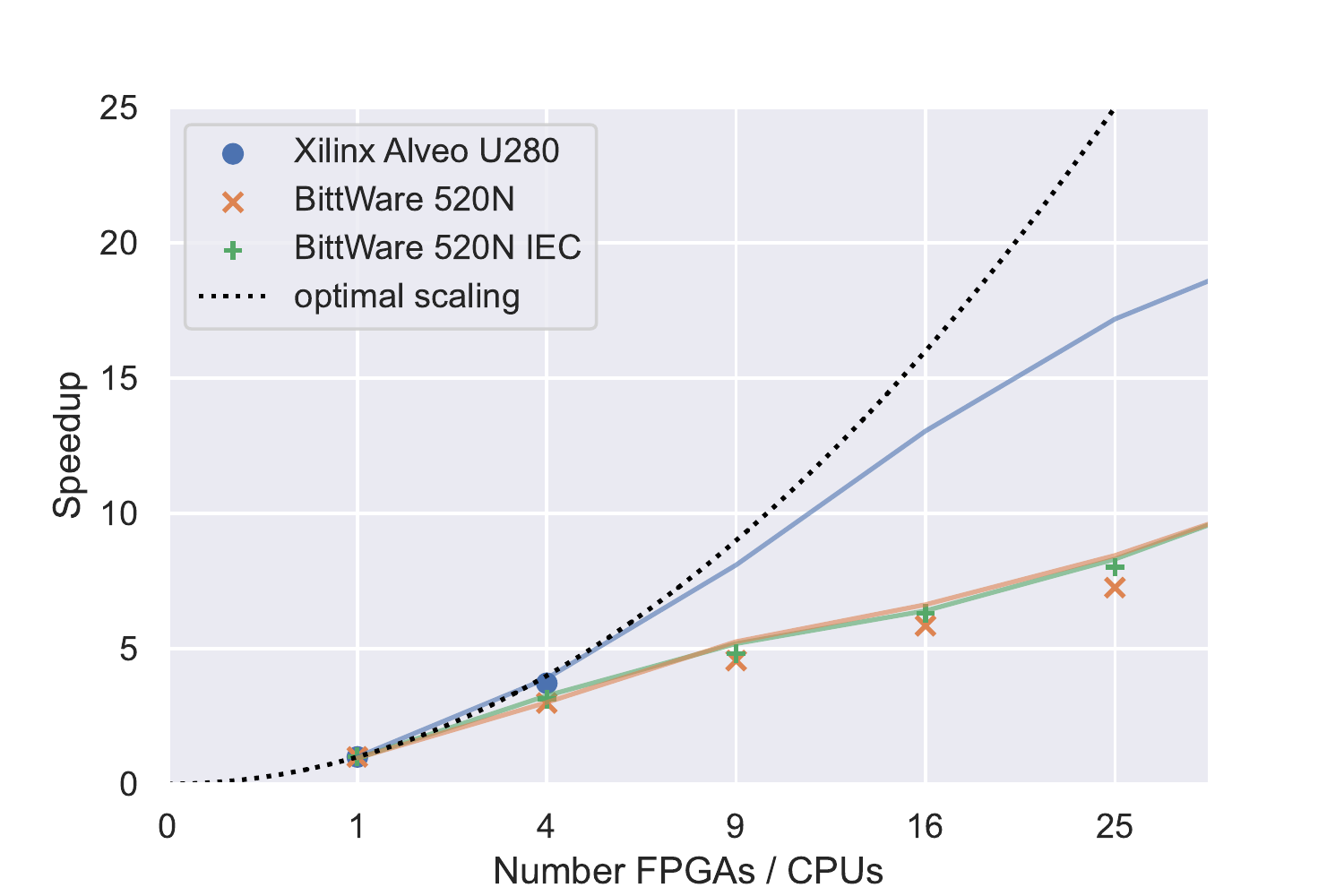}
  \caption{Speedup of HPL with a matrix width of 24,576 elements over multiple FPGAs in a strong scaling scenario. Extrapolation models for the three different bitstreams are given as colored lines. They are based on the measured single-FPGA performance per matrix size in Figure~\ref{fig:hpl_scaling_single_fpga}.}
  \label{fig:hpl_torus_speedup_strong}
\end{figure}

The results of the strong scaling experiment are given in Figue~\ref{fig:hpl_torus_speedup_strong}. 
Both benchmarks show a much lower increase in performance for larger torus sizes.
Based on the data of our single FPGA scaling experiment shown in Figure~\ref{fig:hpl_scaling_single_fpga}, we created an extrapolation model for the strong scaling experiment.
It shows that performance per FPGA is tighlty coupled to the size of the local matrices on the \acp{FPGA}.
The extrapolation for the Xilinx Alveo U280 shows a better speedup in this strong scaling scenario because of the smaller block sizes.
With this very simple approach it is already possible to model the performance depending on the total matrix size and the number of \acp{FPGA} with high accuracy.
The strong scaling experiment shows that the overall performance in the torus is tightly coupled to the input size on a single device for all implementations.

The HPL implementation achieves a lower performance per FPGA than the existing GEMM benchmark, although both get their performance from matrix multiplication.
Also, the configuration parameters are chosen similarly for both benchmarks, which results in a similar expected performance.
The main difference in performance is caused by the different clock frequencies of the designs given in Table~\ref{tab:resource_usage}.
The tools achieve higher clock frequencies for the base implementation of the GEMM benchmark because the kernels of the HPL communication phase consume additional resources.
Also, the matrix multiplications work on smaller matrix sizes of just a single block, which reduces the reuse of data in local memory.

Our HPL implementation achieves 14.3~TFLOP/s for the base version and 20.8~TFLOP/s for the optimized version using \ac{IEC} on 25 BittWare 520N \acp{FPGA}.
The scaling experiments show, that the two major reasons for the performance differences rely on the achieved frequencies and a more efficient use of the global memory.
Although the benchmark is computation-bound, our optimized version using \ac{IEC} still achieves higher performance by reducing the number of \acp{LSU}.
This allows further global memory optimizations and higher kernel frequencies that improve the performance.

\subsection{Evaluation of the Existing Benchmarks}
\label{sec:existing_bms_eval}

For STREAM, FFT, and GEMM, the design did not change compared to the previous work.
All except RandomAccess are executed embarrassingly parallel, so MPI is only used to exchange measurement and validation results.
In the case of RandomAccess, the data array is distributed among the \acp{FPGA}.
This way, only scaling to a power of two is allowed since the total size of the data array must be a power of two.

\begin{figure*}
  \begin{subfigure}{0.49\linewidth}
    \centering
    \includegraphics[width=\linewidth]{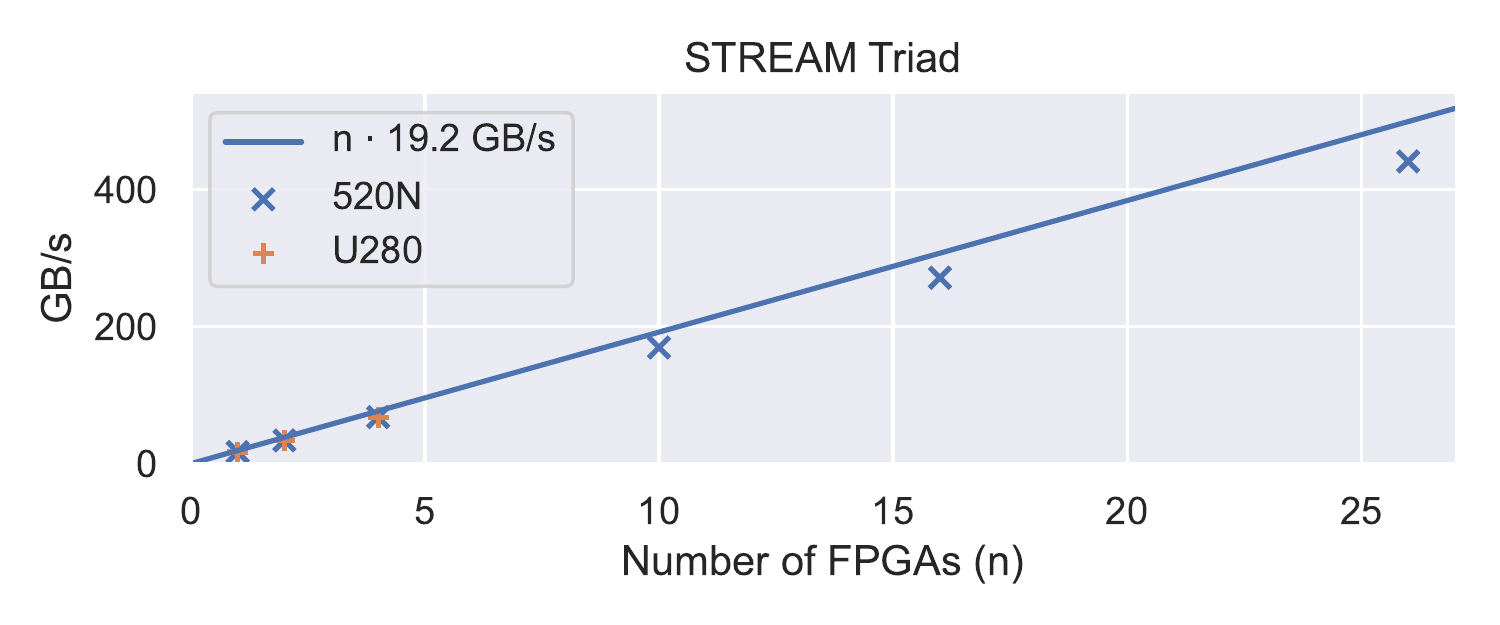}
    \caption{STREAM Triad with 4 GB arrays per FPGA}
    \label{fig:stream_scaling}
  \end{subfigure}
  \begin{subfigure}{0.49\linewidth}
    \centering
    \includegraphics[width=\linewidth]{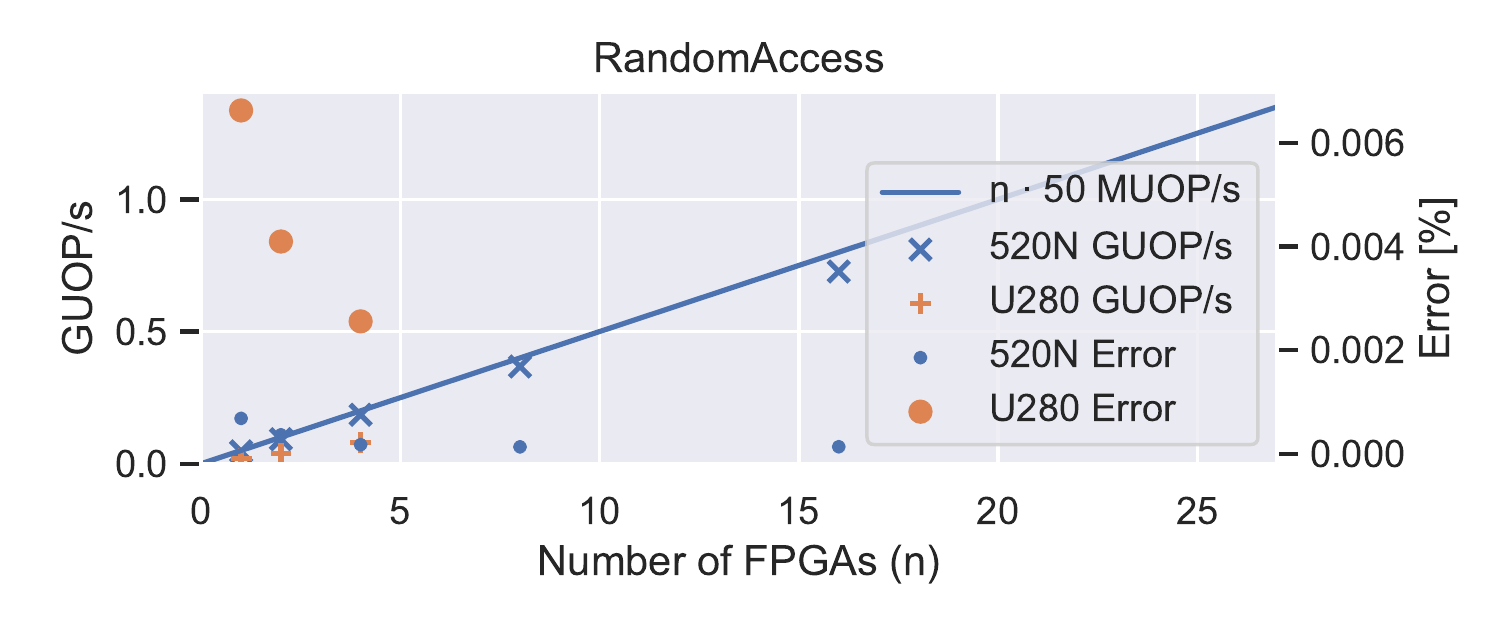}
    \caption{RandomAccess with 8 GB of data per FPGA}
    \label{fig:ra_scaling}
  \end{subfigure}
  \begin{subfigure}{0.49\linewidth}
    \centering
    \includegraphics[width=\linewidth]{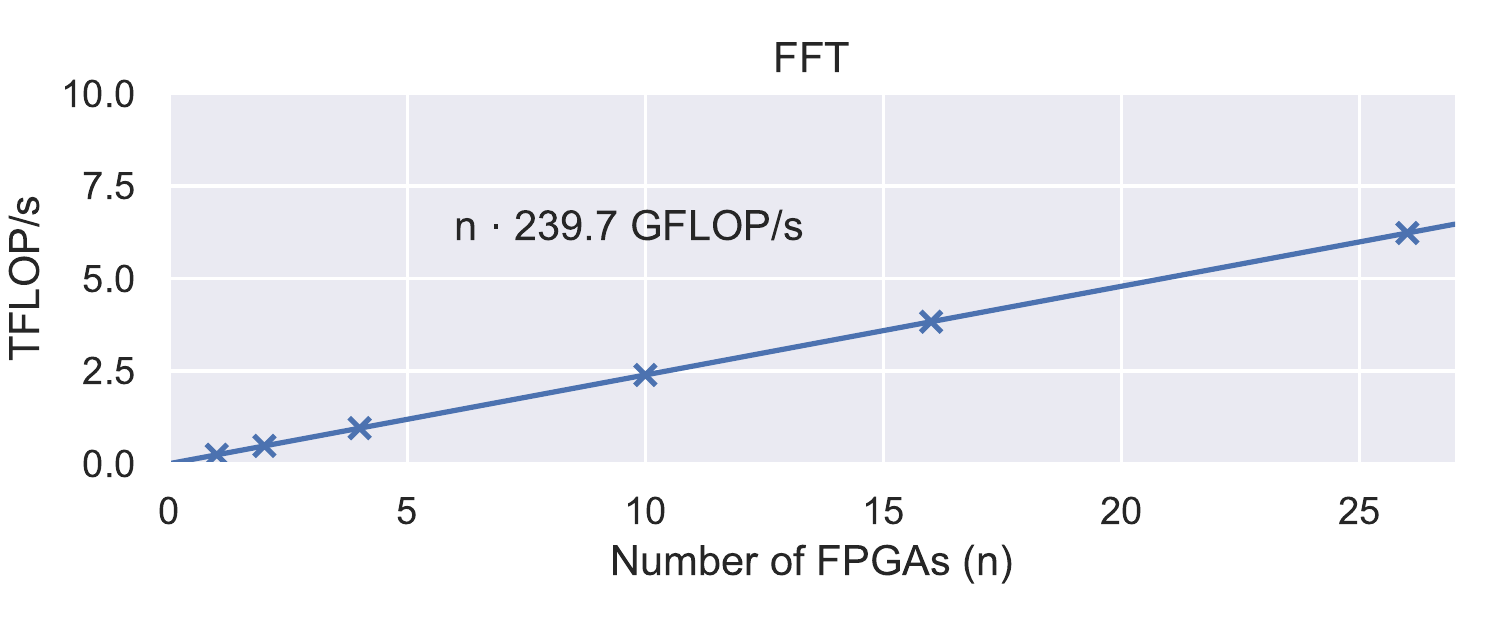}
    \caption{Batches of 4096 1d FFTs of size $2^{17}$ per FPGA}
    \label{fig:fft_scaling}
  \end{subfigure}
  \begin{subfigure}{0.49\linewidth}
    \centering
    \includegraphics[width=\linewidth]{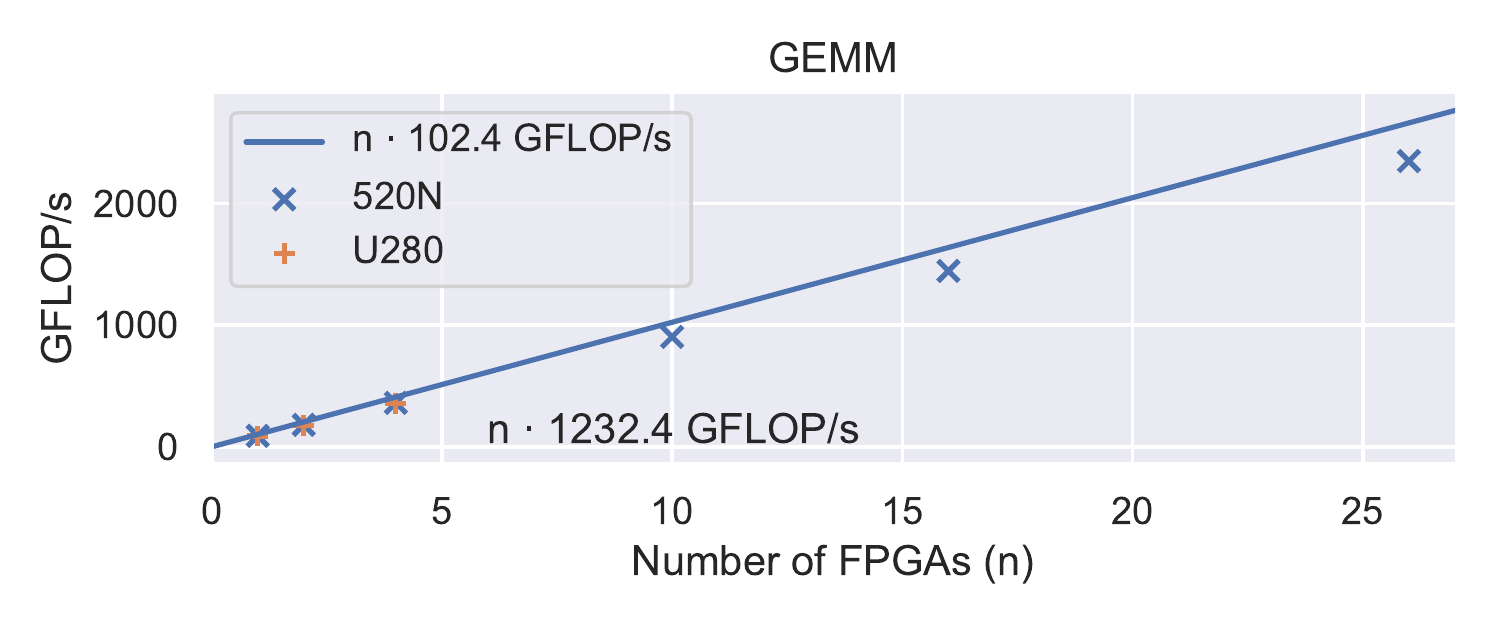}
    \caption{GEMM with a matrix width of 45 blocks per FPGA}
    \label{fig:gemm_scaling}
  \end{subfigure}
  \caption{Normalized performance of the four benchmarks without inter-FPGA communication. Data is normalized to a single memory bank and a frequency of 100~MHz clock for STREAM, GEMM, and RandomAccess, to allow a better comparison for the two FPGA boards.}
  \label{fig:basic_results}
  \end{figure*}

We executed the four benchmarks on up to 26 \acp{FPGA} to show their scaling performance.
4~GB arrays are used in STREAM, FFT calculates on $4,096$ 1d FFTs of $2^{17}$ or $2^9$ complex numbers, and
GEMM uses matrices with the width of $23,040$ elements per \ac{FPGA}.
RandomAccess is executed in a strong scaling scenario with 8~GB data array.
The normalized measurement results are given in Figure~\ref{fig:basic_results}.
For STREAM, the measurements are normalized to a single memory bank with a theoretical bandwidth of 19.2~GB/s.
The benchmark shows a similar scaling behavior on both devices.
For GEMM, the results are normalized to a single kernel replication running at 100~MHz with with an $8 \times 8 \times 8$ matrix multiplication in registers.
This leads to a maximum theoretical performance of 102.4~GFLOP/s times the number of used \acp{FPGA}. Also here, the base implementation shows a performance close to the theoretical peak on both devices.
Because of the comarably high clock frequency of our synthesized design, we achieved more than 1.2~TFLOP/s per FPGA on the BittWare 520N.

Although we did not change the benchmark code, we were not able to execute the FFT benchmark on the Alveo U280.
The benchmark required internal channels or pipes between the kernels to forward data but support for pipes in OpenCL kernel code was removed with XRT 2.9.
This still allows synthesis of the benchmark, but no execution.
On the BittWare 520N, the benchmark scaled lineraly. For FFT, we show the absolute measured performance.

The RandomAccess results were also normalized to the number of memory banks and a kernel frequency of 100~MHz.
Because an update of a value requires one read and one write to the memory bank, two clock cycles are required per update, which results in a theoretical peak performance of 50~MUOP/s per FPGA.
On the BittWare 520N, the base implementation gets close to this theoretical peak whereas on the Alveo U280, we only get roughly half of this performance.
One reason for that is the difference in the configuration: For the Alveo U280 we need a small buffer to read and write multiple values subsequentially which partially hides the lantency of memory accesses and increases performance.
As a trade-off, we see an increased error rate, because we may overwrite values, that are already in the buffer.
Still, this approach requires iteration between two different pipelines that fill and empty the buffer. Since these pipelines have a considerable latency because of memory accesses, this will also reduce the performance because the pipelines have to be emptied frequently.
If it would be possible to ignore the dependency between reads and writes, the single-pipeline approach could be used as it is done for the BittWare 520N.

\section{Related Work}
\label{sec:related_work}

Several benchmark suites for \ac{FPGA} provide benchmarks for OpenCL or \ac{HLS} frameworks \cite{RodiniaFPGA, SHOC, Rosetta, CHO}.
However, the benchmarks often use small input sizes or come with fixed kernel designs that allow no easy scaling to larger sizes.
With Spector \cite{Spector}, a benchmark suite exists that provides configuration parameters for each benchmark to support design space explorations.
Also HPCC FPGA \cite{hpcc_fpga} comes with parametrizable benchmarks and is based on the HPC Challenge benchmark suite \cite{HPCCIntroduction} for CPU, so the benchmarks target the HPC domain.
But it lacks implementations for some of the benchmarks.
None of the suites supports the performance characterization of a multi-\ac{FPGA} system and its inter-\ac{FPGA} communication networks.

The most complex benchmark that we added to the benchmark suite is LINPACK, where the most compute-intensive part is the LU decomposition.
There already exist several implementations for scalable, double-precision blocked LU decomposition implemented in \ac{HDL}.
A multi-FPGA implementation of LU decomposition is proposed in \cite{lu_factorization_multifpga}.
The implementation uses up to five Virtex-II FPGAs for the calculation on a single matrix arranged in a star topology. 
The FPGAs need to be reconfigured several times during computation and calculate the LU decomposition of an $8192 \times 8192$ matrix with double precision complex numbers using 5 FPGAs in 1862.41 seconds.
Wu et al. \cite{claim_multi_fpga} propose a single FPGA implementation for Virtex-5 FPGAs that reaches 8.5 GFLOP/s and that can be easily extended for multi-FPGA execution.
Jaiswal and Chandrachoodan also propose a scalable double-precision block LU decomposition implementation for Virtex-5 FPGAs \cite{scalable_block_lu} written in Verilog.
They report achieving more than 120 GFLOP/s when scaling over 8 FPGAs.

Turkington et al. \cite{linpack_handel_c} implement the LINPACK 1000 benchmark in Handel-C and achieve more than 2.5 GFLOP/s on Stratix II.
Wu et al. \cite{linpack_wu} achieved more than 3.6 GFLOP/s with their \ac{HDL} implementation of the same benchmark.
Both implementations also include pivoting.
Since we are rather taking up on the rules proposed for HPL-AI, pivoting is not part of our implementation.

Zohouri et. al. \cite{linpack_zohouri} implemented an OpenCL single-precision LU decomposition without pivoting within the Rodinia FPGA benchmark suite.
Execution on an Intel Arria 10 FPGA with a matrix size of 8,192 elements resulted in a performance of above 366.5 GFLOP/s.
Our implementation requires much larger matrix sizes to achieve its peak performance but already achieves 493.7 GFLOP/s on Stratix 10 with the given matrix size.
However, since our implementation is also scalable over multiple FPGAs, the overall performance is not limited by the resources of a single FPGA.

\section{Conclusion} 
\label{sec:conclusion}
In this work, we extended the HPCC FPGA benchmark suite with support for multi-FPGA systems and their inter-\ac{FPGA} communication interfaces.
Therefore, we proposed a scalable version of the RandomAccess benchmark and extended all existing benchmarks with multi-\ac{FPGA} support.
Moreover, we added three new benchmarks, b\_eff, PTRANS, and LINPACK, for Xilinx and Intel FPGAs that stress inter-\ac{FPGA} communication and provided baseline implementations via \ac{MPI} and PCIe for all of them.
The baseline implementations show similar normalized performance and scaling behavior on our two evaluation systems with up to 26 BittWare 520N and four Xilinx Alveo U280 boards.

To show the extendability of the benchmark suite with support for vendor-specific communication interfaces, we also provided implementations with \ac{IEC} for direct point-to-point connections between \acp{FPGA}.
Evaluation of the vendor-specific and the baseline implementations revealed the advantages of direct inter-\ac{FPGA} communication over communication via \ac{MPI} and the CPU network not only for the communication-bandwidth-bound applications but also for computation-bound applications like LINPACK.

With LINPACK, we also proposed a well-scaling LU decomposition implementation in a 2D torus.
The evaluation showed that the performance of the implementation is limited by the aggregated matrix multiplication performance of the used devices.
With further architecture-specific optimizations to increase the clock frequency of the implementation, more than 1 TFLOP/s per FPGA on the BittWare 520N are within reach with the proposed design.

We make the extended version of HPCC FPGA publicly available to facilitate active participation in development towards a performance characterization tool for HPC multi-FPGA systems and their inter-FPGA communication interfaces.

\ifopen
\begin{acks}
The authors gratefully acknowledge the support of this project by computing time provided by the Paderborn Center for Parallel Computing (PC2) and the Systems Group at ETH Zurich as well as the Xilinx Adaptive Compute Clusters (XACC) program for access to their Xilinx FPGA evaluation system.

This work is partially funded by the German Research Foundation (DFG) within the Collaborative Research Center "On-The-Fly Computing" (SFB901), the project "Performance and Efficiency in HPC with Custom Computing" (PerficienCC), grant agreement No PL 595/2-1, and by the Federal Ministry of Education and Research (BMBF) and the state of North Rhine-Westphalia as part of the NHR Program.
\end{acks}
\fi

\newpage
\bibliographystyle{bibliography/ACM-Reference-Format}
\bibliography{bibliography/meyer21_sc,bibliography/acmart}

\end{document}